# The role of self-coherence in correlations of bosons and fermions in linear counting experiments.
## *Notes on the wave-particle duality*

### Sándor Varró

*Research Institute for Solid State Physics and Optics*
*of the Hungarian Academy of Sciences*
*H-1525 Budapest, P. O. Box 49, Hungary,*
E-mail: varro@mail.kfki.hu

**Abstract.** Correlations of detection events in two detectors are studied in case of linear excitations of the measuring apparatus. On the basis of classical probability theory and fundamental conservation laws, a general formula is derived for the two-point correlation functions for both bosons and fermions. The results obtained coincide with that derivable from quantum theory which uses quantized field amplitudes. By applying both the particle and the wave picture at the same time, the phenomena of photon bunching and antibunching, photon anticorrelation and fermion antibunching measured in beam experiments are interpreted in the frame of an intuitively clear description.

## 1. Introduction

It is a well-established experimental fact that the basic constituents of radiation and matter have both wave and particle characteristics. An uncountable number of experiments have demonstrated what Dirac [1] has written in his fundamental book on quantum mechanics: "Thus all particles can be made to exhibit interference effects and all wave motion has its energy in the form of quanta." It seems to us, however, that the very content of this statement has been and is still being often forgotten – or simply disregarded – in interpretations of experimental results where the dual nature of quanta in particular manifests itself. As illustrations to this statement, good examples offer themselves, for instance, concerning the various interpretations of the observation, made by Hanbury Brown and Twiss [2] in 1956, of correlations in the fluctuations of photoelectric currents induced by partially coherent light in two detectors. The effect was first observed with radio waves, and made it possible to determine the angular size of stars from the correlated current fluctuations in the detectors. This was possible, because the current fluctuations in a square-law detector are connected to the photon number fluctuations, which are not sensitive to first order coherence properties of the radiation. Subsequently the similar effect was demonstrated in the laboratory with light; the signals of two detectors placed at the opposite sides of a beam splitter showed positive correlation. In contrast to the particle picture for the photons, suggested by the title of their paper [2], the discoverers themselves used a (quantitatively satisfactory) semiclassical interpretation of their results. On the other hand, according to Glauber [3], "The intensity interferometry of Hanbury Brown and Twiss was indeed exploiting the interference of pairs of photons. But they were not completely persuaded, and I am not sure they were ever fully persuaded. … Some nonsensical things were initially said about the interpretation of this experiment, but a correct interpetation was soon given by Purcell [[4]]. … But there were other papers published that still misinterpreted the effect." Though the effect can be treated classically [5-6], the interpretation is usually based on the quantum theory of light [7], on which, in the context of quantum optics, many standard texts [8-11] have been available in the meantime. The measurement of two-particle boson and fermion correlations in various processes has become a useful experimental tool in particle physics, too [12-13]. At this point we would like note that already in the first half of the sixties of the last century, in a series of papers Goldberger, Lewis and Watson [14-18] presented a unified, thorough and very detailed theoretical analysis of this kind of correlation effects in the framework of quantum mechanics, including the description of various sorts of sources, scatterers, detection mechanisms, and giving, moreover, estimates of the all-over important signal to noise ratios. We think that their work does not receive that attention nowadays as it deserves.

Recently there has been a renoved interest in the investigation of correlations of Hanbury Brown and Twiss (HBT) type between both bosons [19-25] and fermions [26-32], and there are various interpretations, even misconceptions showing up from time to time. The reason for that is at least twofold. Roughly speaking, in the quantum description of the original HBT effect the intensity-intensity correlation function $G_{12}^{(2)}$ can be expressed in terms of the amplitude correlation $G_{12}^{(1)}$. This means that $G_{12}^{(2)} = I_1 I_2 + |G_{12}^{(1)}|^2$, where $I_1$ and $I_2$ are the average intensities around the two spatio-temporal centers of the detectors, i.e. the effect can be viewed as a self-coherence effect, which can be explained in terms of stochastic classical fields. Namely, the correlations in the two detectors placed at the opposite sides of a beam splitter are caused by the noise in the source which is mediated by the two component of the split beam. On the other hand, this effect has been considered as a result of quantum interference between two emission processes and two absorption processes taking place at the source and at the two separated detector points, respectively. This kind of process is the photoelectric mixing, which was first theoretically analysed by Fano [33], who described it as a result of quantum interference between the emission and detection of two photons, stemming from two atoms $a$ and $b$ in an extended source, and detected at $c$ and $d$. This 'histories' are represented by the joint transition amplitudes $(a \rightarrow c)(b \rightarrow d)$ and $(a \rightarrow d)(b \rightarrow c)$, which 'cross each other' if one imagines the photons propagating in two directions with some average wave vectors. However, one should keep in mind, that Fano, before closing his paper, felt it necessary to emphasize the following important point: "Averaging over mutual positions of many different pairs yields a nonvanishing effect of constructive interference only for positions so confined that $r_{cd} r_{ba} / \overline{\lambda} R \lesssim 1$. [ Eq. (34) ]. Furthermore, the whole calculation pertains only to such pairs of source atoms that had been simultaneously in an excited state, i.e., which had become excited within a time interval of the order of $1/\overline{\Gamma}_{ab}$." Here $r_{cd}$ and $r_{ba}$ are the distances of the two detector atoms and the source atoms, respectively. $R$ is the mean distance of the sources and the detectors. Moreover, $\overline{\lambda}$ and $1/\overline{\Gamma}_{ab}$ are the mean wavelength of the radiation and the mean life time of the light emitting atoms, respectively. In fact, Fano's above condition is equivalent to the requirement that the whole interaction region should cover possibly the smallest number of coherence volumes $A_c \overline{\lambda}$, in order to observe the effect, where $A_c$ is the coherence area of the radiation at the detection surface. This condition is completely equivalent with the requirement that the number of relevant (but not plane wave) modes of the waves should possibly be a minimum, namely 1. The two sources just pump one mode coherently, and the energy occupying this mode carries the noise to the two detection points and absorbed at random, causing excess correlations between the numbers of their counts. In this sence, this would be a single-photon 'self-coherence' effect. Let us note at this point that in many important cases the 'single-quantumness' of detection events is naturally satisfied, like in the everyday practice of neutron interferometry [34], as has been emphasized by Rauch et al. [35]: "All the performed experiments belong to the regime of self-interference because the phase-space density of any neutron beam is extremely low ($10^{-14}$) and nearly every case when a neutron passes through the interferometer the next neutron is still in a uranium nucleus of the reactor fuel."

The main purpose of the present paper is to derive the correlation statistics, measured in HBT type experiments, in a clean way by keeping track of the simple algebra of events on the basis of classical probability theory. We think that our approach is closer to the physical intuition as the more general (but, on the other hand, often quite involved) quantum field theory, and, at the same time, gives identical numbers. The method to be presented here may perhaps be a usable tool for simpler and cleaner interpretations of several important experiments performed nowadays.

In the following, we shall describe the HBT type (two-point) correlations between the number of counts of bosons and fermions under the strict assumption that there may definitely be only one quantum absorbed in the measuring apparatus *during one elementary interaction process*. This means that the number of true interactions at most coincides with the number of incoming quanta. This assumption, of course, does not exclude the possibility that *during one sequence of measurements* (when the detection gate is open) there are more than one quanta in the apparatus. If this would not be the case, we would not be able to explain any excess correlations. We shall describe the distribution of quanta in the available modes (set by the experimenter) on the basis of classical probability theory. The key element in our method is the proper definition and physical interpretation of the relevant degrees of freedom in an experiment. Concerning this point, see also section 2 of the thoroughly written paper by Oxborrow and Sinclair [82]. We shall not discuss cases when the chain of the single-particle interactions is interrupted by particle creations during the detection time, which would completely modify the distribution of the energy among the degrees of freedom.

In our recent study [36] we have worked out a formalism for treating HBT type correlations in single-photon experiments, more precisely, in the extreme case when one degree of freedom is available for the photon (the case of quasi-monochromaticity in one spatial mode with a given polatization) during the detection time. There the emphasize has been put on the proof of the mathematical identity of our final formulae, obtained in the frame of classical probability theory, with the textbook results derived on the basis of second quantization and Glauber's standard quantum coherence functions [37-38]. The term 'classical' here does not refer to the classical continuous (Maxwell) fields. In the present description the operator algebra of the quantized field amplitudes has been got around by using the classical probability calculus for the *random integer number of quanta* absorbed by the detectors. (This means that the present description has nothing to do with the stochastic electrodynamics which operates with continuous field quantities.) In the present paper we give a generalization of our method to treat cases when several spatio-temporal modes can be excited by the quantum. In order to make the paper possibly self-contained, in Section 2, after an introductory discussion of the physical meaning of our basic assumptions, we briefly summarize the relevant formal results presented already in [36], and derive the basic formulas which will be applied in Sections 3 and 4 in the analysis of some illustrative examples. Section 5 will be devoted to the discussion of the recent HBT type experiments with atoms and neutrons performed by Jeltes *et al.* [25] and Iannuzzi *et al.* [31], respectively. In Section 6 we analyse the fundamental experiments by Aspect and Grangier [58] in terms of our new method. In Section 7 our main assumptions and results shall be summarized, and some general conclusions will be drawn concerning the interpretations of HBT type experiments in the context of wave-particle duality.

## 2. Sequences of elementary measurement acts. Elementary correlations

The single-quantum experiments have been defined in [36] so that during one elementary measurement act the energy $\varepsilon$ of the incoming quantum, available for the absorbers in the measuring apparatus, is enough to excite at most one of the absorbers belonging to the two detectors. In addition we have taciltly assumed linearity, i.e. an absorber cannot be excited by two quanta, or, in other words, there is no 'particle clumping' in the elementary interaction processes. If the spatio-temporal separation of the detectors is reduced, and the two detectors gradually coalesce, then we consider only cases when at most two incoherent quanta (with the same energy but orthogonal polarizations or spin directions) are absorbed, but each of them excite two distinguishable absorbers independently. Thus, in the ideal case, when the whole interaction region is limited to one mode (more precisely one degree of freedom with respect to spatio-temporal modes and polarization or spin), then the measured coincidence rate may deviate from the background ('accidental') coincidence rate. Since the ideal thermal beams contain the 'maximum correlations or disorder in themselves', we expect an increase by a factor of 2 for bosons and a complete supression for fermions, if the mismatch of the spatio-temporal regions of the two detectors are infinitesimally small. This is the reason for why with thermal boson beams this fundamental 'bunching' (quantified by the 'particle clumping' term in the coincidence rates) is always present. Oppositely, we expect an intrinsic tendency for 'anti-bunching' of the counts with thermal fermion beams. The characteristics of the correlations in the measured count distribution do, of course, crucially depend on the nature of the source(s), the physical properties of the detectors and the sampling method [35], i.e. the way how the data acquisition is taking place, for instance by changing the gate duration or by using some triggering mechanism.

According to our recent study [36], we consider an *elementary measurement act* (in other words, an *elementary experiment*) as a *ternary process* (rather than a binary process) in which either detector A, or detector B is excited, or neither of them. These *mutually exclusive* but *not independent* events, $A$, $B$ and $C$ form a complete set, *each with positive probabilities*, $p$, $q$ and $r$, respectively.

$$P(A) = p, \quad P(B) = q, \quad P(C) = r, \quad \text{where} \quad p + q + r = 1, \tag{1a}$$

                                           8

$$A \cap B = O, \quad A \cap C = O, \quad B \cap C = O, \quad A \cup B \cup C = I, \tag{1b}$$

$$0 = P(A \cap B) \neq P(A)P(B) = pq \neq 0, \quad P(A \cap C) \neq P(A)P(C), \quad P(B \cap C) \neq P(B)P(C). \tag{1c}$$

For instance the first relation in Equation (1c) shows that $A$ and $B$ are not independent, because the probability of their product (of their 'intersection', which is the impossible event) is zero, on the other hand the product of their probabilities, $pq$, is clearly nonzero. In Figure 1 we show a symbolic sketch to illustrate the possible physical background of the above events.

SPACE FOR FIGURE 1

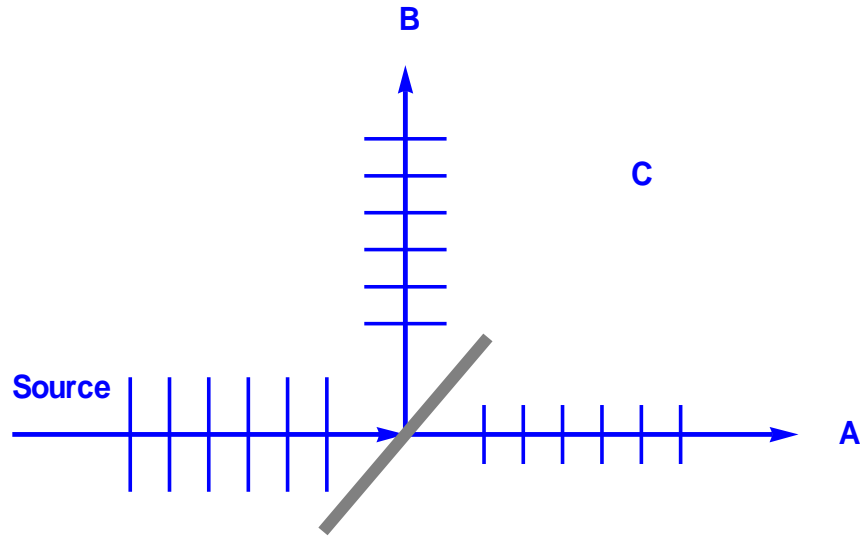

**Figure 1.** Illustration of the three possible outcomes $A$, $B$ and $C$ of an elementary measurement act which are mutually exclusive but not independent. If $P(C) = r = 0$, then there is a strict binomial anticorrelation between $A$ and $B$. In this case the incoming energy would flow into chanel A or B with absolute certainty. In our formulation of the problem, we relax this dichotomy to a trichotomy. This is the key element in the algebraic part of the present analysis, which make our method basically different from that used earlier in treating count distributions in terms of classical binomial probabilities, like in [40] or in [44]. As will be explicitely shown in the main text, event $C$ is compatible with the microscopic conservation of energy, thus, it does not by no means correspond to some kind of dissipation. Concerning the analysis of the beam splitter problem in terms of quantized amplitudes we refer the reader to the works [41-44]. We also note that the source is not considered as a part of the measuring apparatus. Event $C$ is possible, for instance, if some of the quanta are reflected from the entrance port in front of the beam splitter [30]. In some cases the branching point at the middle of our 'symbolic beam splitter' can be consideted as a scatterer, or as an effective source, which elastically redistributes the enrgy into several spatial degrees of freedom, as has been accurately discussed by Goldberger and Watson [16, 17].

To characterize the outcome of a *sequence* of $n$ single-quantum *independent elementary expriments* we introduce the random variable $\xi_n(A)$ being the number of interactions (from altogether a *definite number n* of elementary experiments) when detector A is excited (i.e. the removal of the quantum takes place at detector A). Similarly, the variable $\eta_n(B)$ is the number of independent elementary experiments (from altogether $n$ experiments) in which detector B is excited. The joint distribution of these random variables is the following *trinomial distribution* [36]

$$w_{mk}(n) \equiv P(\xi_n = m, \eta_n = k) = \frac{n!}{m!k!(n-m-k)!} p^m q^k r^{n-m-k}, \qquad (2)$$

where we have taken into account that the order of the results *A*, *B* and *C* within the sequence is immaterial. In general, the calculations of expectation values and higher moments of probability distributions can be conveniently done by using the *generating functions* [36], which we now introduce. The two-variable generating function of the joint distribution given by Equation (2) reads (according to the trinomial expansion of the $n^{th}$ power of the sum of three numbers)

$$G_n(x,y) \equiv \sum_{m=0}^{n} \sum_{k=0}^{n} w_{mk}(n) x^m y^k = (px + qy + r)^n, \qquad (3)$$

where $x$ and $y$ are in general complex subsidiary variables satisfying the relations $|x| \leq 1$ and $|y| \leq 1$. The expectation values and higher moments of the random variables $\xi_n$ and $\eta_n$ can be expressed in terms of the first order and higher order partial derivatives of the generating function, respectively. For instance

$$\overline{\xi}_n \equiv \sum_{m=0}^{n} \sum_{k=0}^{n} m w_{mk}(n) = \left[\frac{\partial G_n(x,y)}{\partial x}\right]_{x=1, y=1} = np, \qquad \overline{\xi_n^2} = np + n(n-1)p^2, \qquad \overline{\eta}_n = nq, \qquad (4)$$

where the upper dash denotes expectation value within a sequence. The second moment of $\eta_n$ has the same form as that of $\xi_n$ with $p$ replaced by $q$. The dispersions (standard deviations) $\Delta\xi_n$ and $\Delta\eta_n$ are defined as the positive square roots of the corresponding variances,

$$\Delta\xi_n^2 \equiv \overline{(\Delta\xi_n)^2} \equiv \overline{(\xi_n - \overline{\xi}_n)^2} = \overline{\xi_n^2} - \overline{\xi}_n^2 = np(1-p), \quad \Delta\eta_n^2 = nq(1-q), \quad \overline{\xi \cdot \eta} = pqn(n-1). \quad (5)$$

The (microscopic) correlation $K_n$ (within a sequence of a definite number of measurement acts) of counting events in detectors A and B are, on one hand, quantified by the expectation value of the product of the number of counts (coincidences within a sequence) relative to the background or 'accidental' coincidences. This quantity is simply related to the mean of the relative fluctuations of the signals,

$$\frac{\overline{(\Delta\xi_n) \cdot (\Delta\eta_n)}}{(\overline{\xi}_n) \cdot (\overline{\eta}_n)} = K_n - 1, \quad K_n \equiv \frac{\overline{\xi_n \cdot \eta_n}}{\overline{\xi}_n \cdot \overline{\eta}_n} = 1 - \frac{1}{n} < 1, \quad \left[ \frac{\langle \hat{a}^+ \hat{a}^+ \hat{a} \hat{a} \rangle_n}{\langle \hat{a}^+ \hat{a} \rangle_n \langle \hat{a}^+ \hat{a} \rangle_n} = 1 - \frac{1}{n} \right]. \quad (6)$$

Equation (6) shows that in an *n–sequence* there is always a *negative correlation* between the detection events $A$ and $B$, regardless of the Bose or Fermi character of the detected quanta. The second equation of Equation (6) may be compared with the normalized intensity-intensity autocorrelation functions of a single-mode photon field in a photon number eigenstate $|n\rangle$ at exactly zero delay [7], which is displayed in the bracket in the third equation. We have denoted by $\hat{a}$ and $\hat{a}^+$ the annihilation and creation operators of the excitation of a single (spatio-temporal) mode, respectively. The combination $\hat{a}^+ \hat{a}^+ \hat{a} \hat{a} = \hat{a}^+ \hat{a} \hat{a}^+ \hat{a} - \hat{a}^+ \hat{a}$ in the bracket stems from the expectation value of the normally ordered product $E^{(-)}(\mathbf{r}_1, t) E^{(-)}(\mathbf{r}_2, t+\tau) E^{(+)}(\mathbf{r}_2, t+\tau) E^{(+)}(\mathbf{r}_1, t)$, where $E^{(+)}$ and $E^{(-)}$ are the positive and negative frequency parts of the relevant electric field components, respectively. If the two detectors are placed within one transverse mode (i.e. $|\mathbf{r}_2 - \mathbf{r}_1|^2 < A_c$, where $A_c$ is the transverse coherence area of the incoming beam), then in the limit $\tau \to 0$ the above fourth-order product is proportional with the combination shown in the bracket.

The appearent similarity does not mean that an *n*-sequence is a representative of a number eigenstate, since the latter one is exactly dispersionfree, in contrast to $\xi_n$ and $\eta_n$. In Equation (6) the integer *n* means the number of possible elementary interactions (the maximum number of the incoming quanta) taking place within the whole detection time (during which the detector's gate is open).

The expectation values $\overline{\xi}_n = np$ and $\overline{\eta}_n = nq$ in Equation (4) are considered by us as microscopic probabilities stemming from an *n*-sequence. In our formalism these calculated probabilities are considered as 'ordinary probabilities' on Borel sets ($\sigma$-algebras) events associated to real experiments. The measured values of the physical quantities are theoretically estimated by calculating ensemle averages and higher moments defined on these Borel sets. In Equation (5) the microscopic expectation value $\overline{\xi \cdot \eta} = pqn(n-1)$ is considered by us as the probability of true coincidences within an *n*-sequence, i.e. within a macroscopic gate. When we use linear detectors, true coincidences can never occur due to the conservation of energy, i.e. at a given sharp instant of time at most only one quantum may be detected. Because in reality no 100% pumping of any finite entrance port is possible, during the excitation of the apparatus there is always a finite probability that neither of the two detectors are excited (which outcome is represented by event *C*). Rather, there is always anticorrelation between the counts. However, within a macroscopic gate we can of course observe joint counts, or various anticorrelation effects, if the resolution of the measuring apparatus allows us to measure such effects. It is important to keep in mind that in this formalism in general, the macroscopic ensemble averages are not defined within a sequence. The size *n* of a sequence can be considered as a random variable in the usual sense, i.e. it can be associated to a Borel set of ordinary events. The definition of the proper $\sigma$-algebra is based on taking quantitatively into account the macroscopic boundary conditions.

In a series of macroscopic measurements, during a *whole experimental run* the number of gates must possible be very large. These whole experimental runs must be repeated many times, in order to have an acceptable signal to noise ratio in the *complete experimental run*, which gives the *final result* of an experiment. We have assumed in the present paper, at the outset, that the elementary measurement act is instantaneous, i.e. the removal of the quantum from the interaction region takes practially no time. This assumption is supported, e.g. by the well-known experimental fact that the photoelectrons appear always 'promply' when one illuminates the surface of the photodetector, i.e. the 'accumulation time' is practically zero. It is clear, on the other hand, that an ensemble average over $n$, besides the efficiency, must contain the autocorrelation function of the detectors, too, which contains the very response time, dead time or shaping time.

The second important quantity to be considered is the *correlation coefficient* $R_n$, which may be called *normalized covariance*. It is given in our case as [36]

$$R(\xi_n, \eta_n) \equiv \frac{\overline{(\xi_n - \overline{\xi}_n) \cdot (\eta_n - \overline{\eta}_n)}}{\Delta\xi_n \Delta\eta_n} = \frac{\overline{\xi_n \cdot \eta_n} - \overline{\xi}_n \cdot \overline{\eta}_n}{\Delta\xi_n \Delta\eta_n} = -\sqrt{\frac{pq}{(1-p)(1-q)}}, \quad |R(\xi_n, \eta_n)| \leq 1, \quad (7)$$

and, can be brought to the following more tractable form

$$R(\xi_n, \eta_n) = -\sqrt{\frac{s\widetilde{T} \cdot s\overline{T}}{(1 - s\widetilde{T})(1 - s\overline{T})}}, \quad p \equiv s\widetilde{T}, \quad q \equiv s\overline{T}, \quad s \equiv 1 - r, \qquad (0 < s < 1) \qquad (8)$$

where, according to Equation (1a), $\widetilde{T} + \overline{T} = 1$. It is important to note that if $p = q$, then the correlation coefficient reduces to $-p/(1-p)$, but it cannot reach its minimum value -1 only in the case when $r = 0$, i.e. $p = q = 1/2$ (in which case there is a linear dependence between $\xi_n$ and $\eta_n$). In the symmetric configuration ($\widetilde{T} = \overline{T} = 1/2$), owing to the constraint equation $2p + r = 0$ with positive $r$, the condition $p < 1/2$ has to be satisfied. On the other hand, if $r = 0$, then the trinomial distribution degenerates to binomial distribution of binary alternatives, and $q = 1 - p$, i.e. there is a strict anticorrelation within a sequence between $\xi_n$ and $\eta_n = n - \xi_n$, and the correlation coefficient reaches its lowest possible value $-1$. The normalized correlation of counts $K_n$ depends only on the size $n$ of the sequence, but does not depend on the parameters. On the other hand, the correlation coefficient $R_n$ does not depend on $n$, but it does depend on the parameters. Thus, at least in this sense, the second equations of Equation (6) and Equation (7) are universal, and, moreover they are valid for counting of both bosons and fermions.

**3. Serieses of sequences of elementary measurement acts.**

In reality the number of elementary experiments (i.e. the number of quanta exciting the measuring apparatus during one gate) can never be sharply defined, but rather, it is a random variable with some distribution, which we shall characterize by certain sets of weights $\{W_n, n = 0, 1, 2, ...\}$. Of course, we do not know in which sequence the quanta were detected, hence in this sense we are dealing with a *mixture*. In the present description such a *series of sequences* represent the *whole experimental run*, during which *continuous data acquisition* is taking place. In a series the number of possible outcomes are again two random integers, $\xi(A)$ and $\eta(B)$. The variables $\xi(A)$ and $\eta(B)$ are the number of independent elementary experiments in which detector A or detector B is excited, respectively. At this point we would like again to emphasize that *we are still assuming that during one elementary measurement either* A (*and not* B), *or* B (*and not* A) *or neither* A *nor* B *are excited* (*absorb a quantum*). This is a linearity requirement for the absorbers. The joint distribution of $\xi$ and $\eta$, and the corresponding generating function [36] are now given as the weighted sums

$$P(\xi = m, \eta = k) = \sum_{n=0}^{\infty} W_n P(\xi_n = m, \eta_n = k), \quad G(x, y) = \sum_{n=0}^{\infty} W_n G_n(x, y), \quad \text{with} \quad \sum_{n=0}^{\infty} W_n = 1, \quad (8)$$

where $G_n(x, y)$ has already been defined in Equation (3). In general the weigths $W_n$ depend on the spatio-temporal positions $\{\mathbf{R}_1, T_1, \mathbf{R}_2, T_2\}$ of the detectors ($D$), on the response properties (resolution), and on the statistics of the source of quanta ($S$). The weights also depend on the material constitution (elastic scattering cross-sections, etc.) of the beam splitter ($BS$). These dependencies on all the parameters can be symbolized in short, like this

$$W_n \equiv W_n(M) = W_n[M(\boldsymbol{R_1}, T_1; \boldsymbol{R_2}, T_2 \mid S, BS, D)]. \tag{9}$$

The very meaning of Equation (9) is that the boundary conditions met by the quantum, the response properties of the measuring apparatus, the source and beamsplitter characteristics are condensed into the weigths $W_n$, which are uniquely governed by the number of relevant modes, which we denote by $M$. These weights represent the scene of the scattering experiments, in other words, they set completely the available mode configuration where the energy of the consecutive quanta flowing in and absorbed. Besides the absorption process in the detections, the quantum features come into play in two ways, namely, on one hand, through de Broglie's dispersion relations, connecting energy with frequency, and momentum with wavelength. On the basis of these relations we can calculate the number of relevant modes or, in other words, degrees of freedom. According to Planck's relation, the excitation energy of a single mode (if it is not empty) has a lower limit, namely $h\nu$, and the possible energy content is always an integer multiple of this $h\nu$. This means that if there are more than one potentially active modes available, then the energy of a single quantum necessarily has to flow only into one of them, i.e., there must be an inherent anticorrelation present in such cases. This is the physical content of events $A$ and $B$, which may be called the 'particle face' of the quantum. On the other hand, we should not forget that this does not mean a spatio-temporal localization during a detection gate (exept at the detectors, where, the quantum is removed anyway, and converted to other forms of energy in a microscopic region). Thus, regardless of how large or small the excitation degree of a particular mode is, the interference pattern has already been "encoded" in the true mode function, which takes into account the boundary conditions determined by the *whole* measuring apparatus. This is the manifestation of 'self-interference' in the cases we are discussing. In mathematical terms, the number of modes $M$ can be viewed as functional of stochastic processes and fields characterising the surrounding of the measured quanta. In principle, $M$ is a positive integer, but its value cannot be contolled sharply in many realistic situations. In one of the main types of experiments the source can be represented by some given discrete distributions, and the experimental environment is modified from one whole experiment to an other. Moreover, the spatial modification, i.e. changing $R_1$ and $R_2$, by shifting one of the detectors on the opposite sides of a beam splitter (that way, that the shift is parallel with the propagation of one component of the split beam), a spatial shift is in essence equivalent with a temporal shift. The shift $\Delta x$ can be interrelated to a time delay $\Delta t$ by the relation $\Delta x = \upsilon_g \Delta t$, where $\upsilon_g$ is the *average* group velocity of the quanta. If the beam splitter is an elastic scatterer with practically zero dispersion, then for quasi-monoenergetic beams, this conversion does not cause considerable measurement error. The *width* of the coincidence 'bump' or 'dip', to a good approximation equals $\Delta t / \tau_c \approx M_l \approx \Delta x / \lambda_c$ in this case, where $M_l$ is the number of relevant modes swept in the experiment. The heigth or depth of the excess or missing coincidences (contrast) are determined by the statistical properties of the source. In this case the weigths in Equation (9) can be written in the form $W_n \equiv W_n(M) = W_n[M(M_l, M_{max} | S, BS, D)]$, where in many important cases there is essentially a one-to-one correspondence between the maximum number $M_{max}$ of relevant lateral (transverse) modes and the width of the distribution characterizing the source $S$.

By using Equations (3), (4), (5) and (8) for one spatial mode of the quantum, we can derive the general expressions for the first two moments which are usually used in characterizing the number of counts in a series,

$$\overline{\xi} = p\langle n \rangle, \quad \overline{\xi^2} = p(1-p)\langle n \rangle + p^2 \langle n^2 \rangle, \quad \langle \overline{\Delta \xi^2} \rangle = p(1-p)\langle n \rangle, \quad \overline{\xi \cdot \eta} = pq(\langle n^2 \rangle - \langle n \rangle). \quad (10)$$

Equation (10) is analogous to Equations (4) and (5), but here the bracket refers to expectation values with respect to the weights (ensemble) shown in Equation (9),

$$\langle n \rangle \equiv \sum_{n=0}^{\infty} n W_n, \quad \langle n^2 \rangle \equiv \sum_{n=0}^{\infty} n^2 W_n, \quad K \equiv \frac{\overline{\xi \cdot \eta}}{\overline{\xi} \cdot \overline{\eta}} = 1 + \frac{1}{\langle n \rangle}(F-1) \equiv 1 + \frac{1}{\langle n \rangle}\widetilde{Q}, \quad (11a)$$

$$F \equiv \frac{\langle \Delta n^2 \rangle}{\langle n \rangle}, \quad \widetilde{Q} \equiv F - 1, \quad R(\xi,\eta) = \sqrt{\frac{s\widetilde{T} \cdot s\overline{T}}{(1-s\widetilde{T})(1-s\overline{T})}}(F-1), \quad \widetilde{T} + \overline{T} = 1. \quad (11b)$$

In Equation (11b) we have introduced the Fano factor $F$ [47, 48], which has the same form for both bosons and fermions. In this respect this equation contains quite general formulas with a wide validity. The parameter $\widetilde{Q}$, introduced in the second equation of Equation (11b), is an analogon of Mandel's $Q$ parameter [49], used in quantum optics for characterising photon states and sources. In a single monochromatic spatial mode with a given polarization, the inequivality $\langle n \rangle \leq 1$ always has to be satisfied for fermions, thus $F \leq 1$, and the correlation coefficient is always negative. For photons (or, in general, for bosons) $F$ can either be smaller or larger than one, depending on the occupation statistics of the particular mode, governed by the source. If $F > 1$, i.e $\widetilde{Q} > 0$, then there is a positive correlation between the counts, and if $F < 1$, i.e $\widetilde{Q} < 0$, then there is a negative correlation between the counts. It is interesting to note that the quite general formulas in Equation (11a-b) have been deduced from the simple Boole algebra of the counting events $A$, $B$ and $C$, and from the associated trinomial distribution. We have recovered the well-known result that, depending on the properties of the source (pumping), photons can produce both bunching ($F > 1$) [50], or antibunching ($F < 1$) [51] *in the counting events*. In Ref. [36] we have given the expicite forms of the quantities listed in Equation (11) for the special cases of coherent, thermal, squeezed and phase excitations in a single temporal (quasi-monochromatic) mode. For instance, the coherent excitations described by Poisson weights of the form $c_n = (\overline{n})^n e^{-\overline{n}}/(\overline{n})!$

of arbitrary parameter $\bar{n}$ give $\langle \Delta n^2 \rangle = \bar{n} = \langle n \rangle$, thus the Fano factor is unity, $F = 1$, and the correlation coefficient, $R$ is zero. The relative coincidence rate, $K$ in this case is unity, thus one observes only accidental coincidences. One observes that at detector A the flux of counts is being proportional with $s\widetilde{T}\langle n \rangle$, and one can observe its fluctuations. At detector B one observes the flux of counts proportional with $s\overline{T}\langle n \rangle$, and its independent fluctuations. However higher moments and the moments of the cross-products one measures, one does not notice any correlations. This is, of course, quite naturally expected, because the source follows the 'law of rare events'. Thus, by considering for instance photons, one can conclude that in the measurement the self-coherence of spontaneous emission of a single atom manifest itself. This is possible if one is able to eliminate all the sources of noise (around a particular frequency) in the measuring apparatus within the spontaneous life time of that microscopic transition from which the photon stems.

The numerical identity of our results in Equations (11a-b) with that derived on the basis of second quantization of amplitudes, however, does not mean a conceptual identity with this standard description. Ensemble averages and time averages are sometimes replaced by each other, without any special care. This question belongs to the more general problem of ergodicity, which is, in the context of counting experiments, briefly discussed, e.g. in Ref. [34], where further references can be found. In the present description the averaging (the calculation of physically measurable moments of distributions) is done on the same footing for both the single spatial (quasi-monochromatic) modes (within the basic $n$-sequences and temporal serieses), and for several spatial and other (internal) degrees of freedom (like polarization and spin). In mathematical terminology, this means that we uniformly represent the whole experimental run in the product space of the elementary events $A$, $B$ and $C$ in the $n$-sequences (temporal gates (or windows) in the experiment). It seems that our procedure, which is based e.g. for photons exclusively on semiclassical radiation theory and on classical probability, works quite well for a wide class of linear counting experiments. Moreover, as we shall see below, it can also be applied for treating two-point correlations of any kind of quanta. However, we emphasize that at this level, we consider our approach merely as a more intuitive phenomenological model, in comparison with the 'canonical' scheme of second quantization. We do not attemp to replace the quantum field theoretic description of details of the interactions by using classical and continuous fields.

**4. Ensembles averages with respect to spatial modes and other degrees of freedom**

In the following we illustrate on a couple of important examples how our formalism works in the many mode case. Take first the case of Poisson excitation of the measuring apparatus, by superimposing an arbitrary number of poissonian distributions. The weigths give the probability that exactly $n$ particle occupy $M$ cells (including now the spatial modes, too). The weights are now $M$-fold convolution of the many Poisson distributions, where, in general, we allow arbitrary mean occupations $\bar{n}_1$, $\bar{n}_2$, ... $\bar{n}_M$ for the different (spatial and/or temporal) modes. These $M$-fold convolutions are again Poisson distributions of the form

$$W_n = P_n^{coh}(M) = \sum_{k_1+k_2+...+k_M=n} c_{k_1} c_{k_2} \cdot ... \cdot c_{k_M} = \frac{(\bar{n}M)^n}{(n)!} e^{-\bar{n}M},$$

$$c_{k_i} = \frac{\bar{n}^{k_i}}{(k)!} e^{-\bar{k}_i} \quad (i=1, 2, ..., M), \qquad \bar{n}_1 + \bar{n}_2 + ... + \bar{n}_M \equiv \bar{n}M. \qquad (12)$$

By using the definition of the generating function in Equation (8) we obtain the expression

$$G^{coh}(x, y) = \exp[\bar{n}Mp(x-1)] \cdot \exp[\bar{n}Mq(y-1)], \qquad (13)$$

which shows that in the case of Poisson excitation, the generating function factorizes, thus there are no coincidences, except for the accidental ones (the Fano factor defined in Equation (11) is unity in this case). Moreover, there is a *theorem* in classical probability theory [45], that in the Poisson case the factorization means *perfect independence* (which means much more than merely zero correlation in the second moments) of the number of counts. Thus, all the moments factorize exactly, i. e. $\overline{(\xi^k \cdot \eta^l)} = \overline{\xi^k} \cdot \overline{\eta^l}$ for arbitrary $k \geq 1$ and $l \geq 1$. This corresponds to the case of '*perfect coherence*' in the terminology of Glauber's theory [37, 38] of quantum coherence functions. As has been already mentioned, the absence of any excess correlations in the number of counts is not an unexpected result in this case, if one remembers that it is just the Poisson distribution which describes the fundamental 'law of rare events'. If we can sustain a stationary flow of quanta, by a special source, such that the flow follows the law of rare events, then the self-coherence can perfectly manifest itself. If many source atoms coherently pump a very limited number of modes, then the coherence length is increased, due to constructive interference, and the self-coherence is amplified. The distribution still follows the law of rare events, but now the radiator is a 'macroscopic atom', a large classical dipole. This is the reason for why the laser radiation has Poisson counting statistics. In this case the measuring apparatus can observe only accidental coincidences delivering no other information in the counting regime, than moments of independent fluxes.

Among the sources of radiations, the thermal sources have a special importance, because they show universal features, regardless of the species of quanta they consist of. In case of thermal boson beams, the occupation probabilities can be calculated by a simple combinatorial analysis, because the combinations are equally probable due to complete disorder, yielding the well-known geometric distribution [45],

$$p_k = (1-b)b^k, \quad b \equiv \frac{\overline{k}}{1+\overline{k}} = exp(-\varepsilon/k_B T), \quad (k = 0, 1, 2, ...), \quad \overline{k} = \frac{1}{exp(\varepsilon/k_B T)-1}, \quad (14)$$

where $\varepsilon = h\nu$ is the energy of a quantum, $k_B$ and $T$ are the Boltzmann constant and the absolute temperature, respectively, and $\overline{k}$ is the mean occupation number. The mean occupation number is also termed as degeneracy parameter, whose importance has been emphasized by Mandel [52], already in the early years of photon counting experiments. In a narrow spectral range, i.e. $\bar{n}_1 \approx \bar{n}_2 \approx ... \approx \bar{n}_M$, the probability $B_n^{pol}(M)$ that exactly $n$ identically polarized bosons occupy $M$ spatial modes is given by the $M$-fold convolution, which is a negative binomial distribution of order $M$,

$$B_n^{pol}(M) = \sum_{k_1+k_2+...+k_M=n} p_{k_1} p_{k_2} \cdot ... \cdot p_{k_M} = \binom{M-1+n}{M-1}(1-b)^M b^n, \qquad b = \frac{\bar{n}}{1+\bar{n}}, \tag{15}$$

where $\bar{n}M \equiv \bar{n}_1 + \bar{n}_2 + ... + \bar{n}_M$. More generally, in a thermal beam of arbitrary degree of polarization $P$ ($0 \leq P \leq 1$), the total occupation number of a spatio-temporal mode consist of two statistically independent terms (according to possible polarizations), with mean values $\bar{n}_1$ and $\bar{n}_2$, where $\bar{n}_1 + \bar{n}_2 = \bar{n}$. Then, the probability that exactly $n$ photon occupy $M$ spatial mode (within a narrow spectral range) is the following convolution

$$B_n(M) = \sum_{k=0}^{n} \binom{M-1+k}{M-1}(1-b_1)^M b_1^k \binom{M-1+n-k}{M-1}(1-b_2)^M b_2^{n-k},$$

$$b_{1,2} = \frac{\bar{n}_{1,2}}{1+\bar{n}_{1,2}}, \qquad \bar{n}_1 = \frac{\bar{n}}{2}(1+P), \qquad \bar{n}_2 = \frac{\bar{n}}{2}(1-P). \tag{16}$$

For unpolarized boson beams, the number of relevant degrees of freedom is doubled, $M \to 2 \times M$, because the two possible independent polarization states (as additional 'internal degrees of freedom') count with equal weights. In this case $P=0$, and $\bar{n}_1 = \bar{n}_2 = \bar{n}/2$, i.e. $b_1 = b_2$ in Equation (16), and the summation gives a simple expression, which is again a negative binomial distribution, but now, of order $2M$,

$$B_n^{unpol}(M) = (1-b_1)^{2M}(-b_1)^n \sum_{k=0}^{n} \binom{-M}{k}\binom{-M}{n-k} = \binom{-2M}{n}(1-b_1)^{2M}(-b_1)^n$$

$$= \binom{2M-1+n}{2M-1}(1-b_1)^{2M} b_1^n, \qquad b_1 = \frac{\bar{n}/2}{1+\bar{n}/2} = \frac{\bar{n}}{2+\bar{n}}, \qquad 1-b_1 = \frac{2}{2+\bar{n}}, \tag{17}$$

$$G_N^{boson}(x,y) = \sum_{n=0}^{\infty} B_n(N) = \frac{(1-b)^N}{[1-b(px+qy+r)]^N}.$$

In the last equation of Equation (17) we have summarized the results for the generating functions valid in the two extreme cases of polarized ($N = M$) and unpolarized ($N = 2M$) thermal bosons. From Equations (4), (5), (15) and (17) we obtain for $K_{boson}$

$$K_{boson}^{pol} = \frac{\overline{\xi \cdot \eta}}{\overline{\xi} \cdot \overline{\eta}} = 1 + \frac{1}{M}, \qquad K_{boson}^{unpol} = \frac{\overline{\xi \cdot \eta}}{\overline{\xi} \cdot \overline{\eta}} = 1 + \frac{1}{2M}. \tag{18}$$

These formulas are equivalent expressions for the energy fluctuations of thermal bosons, in general, if the number of modes has a zero dispersion ($< \Delta M^2 > = 0$). This can be seen e.g. by multiplying the variance $\Delta \xi^2$ with the square of the quantum energy $(h\nu)^2$. We obtain $\Delta E_A^2 = h\nu \overline{E}_A + \overline{E}_A^2 / M$, which, in the special case of photons, is just Einstein's fluctuation formula [73] for the energy in a sub-volume in a *Hohlraum* filled with black-body radiation. The first term is called the 'particle-like fluctuation' which would only be present for classical (Poisson) particles. This is the only term present in the extreme Wien limit ($h\nu \gg k_B T$), thus, in this case the radiation behaves so, as if it consisted of Einstein's original 'Lichtquanten' ('light quanta'). The second term comes from the wave character of the photons in the black body radiation, and it is an interferece term of bosons, in general. In this case, the positive correlation between the counting events at A and B are caused by the the self-correlation of the noise compressed to the measurement apparatus in a narrow frequency range and beam cross-section, which is mediated by the flow of quanta. If the two detectors are coalescing within one mode, then the self-coherence perfectly manifests itself, and there is a 'bump' in the coincidence curve with a contrast ratio 2. If the interaction region is increased by separating the detectors more and more, than the number of modes are increasing, thus the contrast $1/M$ goes to zero.

The correlation coefficient given by Equation (11b) can reach its maximum value 1, if the partition of the incoming flow is completely symmetric ($\tilde{T} = \overline{T} = 1/2$). The condition for the maximum value is

$$R^{thermal} = \frac{s\widetilde{T}}{1-s\widetilde{T}}\langle n \rangle = 1, \quad \frac{s}{2} = \frac{1}{1+\langle n \rangle} \equiv 1-b' \equiv 1-e^{-h\nu/k_B T'}, \tag{18a}$$

where $\widetilde{T} = 1/2$, and we have introduced the effective temperature $T'$ associated to the ensemble $\{W_n(M)\}$, which represents the surroundings of the absorption events. According to Equation (18a), since $s < 1$, the average number of detected photons has to satisfy the condition $\langle n \rangle > 1$, if we want to reach the maximum contrast. From Equation (18a), in cases when $s \sim 1$ and $\langle n \rangle \approx 1$, we have $2 \times h\nu \approx 2 \times k_B T' \log 2$. The absorbed two energy quanta are shared as two equal parts of energy $k_B T' \log 2$ to each of the detectors separately. This amount is needed for aquiering the "first bit" of information, according to Gabor [74]. Since we are allowed to assume that the measuring apparatus is already in local thermal equilibrium with the stationary beam, we take $T' \approx T$. The entropy increase of the measuring apparatus is then calculated as a result of an isoterm process, where Planck's expression for the entropy is to be used [74],

$$\Delta S = k_B [(1+\langle n \rangle)\log(1+\langle n \rangle) - \langle n \rangle \log \langle n \rangle] \approx k_B 2\log 2, \tag{18b}$$

where $\Delta S$ is the change of the entropy of a spectral component of the black-body radiation. On the other hand, by writing this same entropy in terms of the energy change and the absolute temperature, we really see, that the information $\Delta H$ aquired by the observer through this elementary coincidence is just the 'first' 1+1 bits in observing a true coincidence of a pair of independent quanta,

$$\Delta S = \Delta E / T \approx k_B 2\log 2, \quad \Delta E = 2h\nu = 2 \times (k_B T \log 2), \quad \Delta H = 2 \times bit. \tag{18c}$$

Now we discuss the case of thermal fermions. Due to the Pauli principle, in thermal equilibrium the probabilities $p_1$ and $p_0$ that an electron (or other fermion) occupies a mode or not, are simply $p_1 = \overline{n}$ and $p_0 = 1 - \overline{n}$, respectively, where $\overline{n} = 1/[\exp(\varepsilon/k_B T)+1]$ is the average occupation number which cannot be larger than 1 (the chemical potential of a beam can certainly taken to be zero). Accordingly, the probability $F_n^{pol}(M)$ that exactly $n$ fermions of identical polarization occupy $M$ spatial modes is given by the binomial distribution of order $M$,

$$F_n^{pol}(M) = \binom{M}{n}(\bar{n})^n(1-\bar{n})^{M-n}, \quad (0 < \bar{n} \leq 1). \tag{19a}$$

In case of a thermal beam of arbitrary degree of polarization $P$ ($0 \leq P \leq 1$), the total occupation number of a mode consist of two statistically independent summands, with mean values $\bar{n}_1$ and $\bar{n}_2$, thus the probability of that exactly $n$ spin–1/2 fermion occupy $M$ spatial mode is the following convolution

$$F_n(M) = \sum_{k=0}^{n}\binom{M}{k}\bar{n}_1^k(1-\bar{n}_1)^{M-k}\binom{M}{n-k}\bar{n}_2^{n-k}(1-\bar{n}_2)^{M-(n-k)},$$

$$\bar{n}_1 = \frac{\bar{n}}{2}(1+P), \ \bar{n}_2 = \frac{\bar{n}}{2}(1-P). \tag{19b}$$

In cases of unpolarized beams we have $P=0$, i.e. $\bar{n}_1 = \bar{n}_2 = \bar{n}/2$, and the weights become

$$F_n^{unpol}(M) = (\bar{n}/2)^n[1-(\bar{n}/2)]^{2M-n}\sum_{k=0}^{n}\binom{M}{k}\binom{M}{n-k}$$
$$= \binom{2M}{n}(\bar{n}/2)^n[1-(\bar{n}/2)]^{2M-n} \tag{19c}$$

$$G_N^{fermion}(x,y) = \sum_{n=0}^{N}F_n(N)G_n(x,y) = [1+\bar{n}(px+qy+r-1)]^N, \tag{19d}$$

where we have summarized the results for the generating functions in the two extreme cases of polarized ($N=M$) and of unpolarized ($N=2M$) thermal fermions.

From Equations (4), (5) and (19a,c) we can easily calculate the necessary moments, by taking the partial derivatives of the generating functions with respect to the subsidiary variables $x$ and $y$, and we obtain for the normalized number of coincidences, $K_{fermion}$, the expressions

$$K_{fermion}^{pol} = \frac{\overline{\xi\cdot\eta}}{\bar{\xi}\cdot\bar{\eta}} = 1-\frac{1}{M}, \quad K_{fermion}^{unpol} = \frac{\overline{\xi\cdot\eta}}{\bar{\xi}\cdot\bar{\eta}} = 1-\frac{1}{2M}. \tag{20}$$

This is an equivalent expression for the energy fluctuations of thermal fermions, which was first presented by Jordan and Wigner in their paper on the second quantization of arbitrary fermion fields. The fluctuation of the energy can be similarly derived, as for bosons, yielding $\Delta E_A^2 = h\nu \overline{E}_A - \overline{E}_A^2 / M$, where the negative sign expresses the 'repulsion' of free fermionic waves.

Equations (18) and (20) express mathematically two of the central results of the present paper. They are in complete accord with the (more general) results published by Goldberger, Lewis and Watson [14-18] on correlations in beam experiments. For photons, Equation (18) coincides with the result obtained from Glauber's standard theory of quantum coherence functions [37-38]. The appearent simplicity of Equations (18) and (20) hide many completely different physical situations through the dependence of the mode function on the experimental conditions. Before entering into the discussion of the physical content of Equations (18) and (20), we note that the weights given by Equations (15), (17) and (19), all go over to classical Poisson weigths in the limit of very large mode numbers and small occupancy, if we keep the product of these quantities fixed,

$$\lim_{M\to\infty,\ \overline{n}\to 0} B_n(M) \ and \ F_n(M) = \frac{<n>^n}{(n)!} exp(-<n>), \qquad <n> \equiv \overline{n}M \ \ fixed. \tag{21}$$

In this case the generating function factorizes, like in Equation (13), and the random variables $\xi$ and $\eta$, representing the number of counts, become independent to any order, thus no excess coincidences are expected in this 'classical case' of rare events.

The number of relevant modes $M$, which is the only variable in our Equations (18) and (20) is determined by the characteristics of the detected particles associated to spatio-temporal changes. *Both in the known quantum descriptions and in the classical descriptions, these functions, in fact, quantify the overlap of the modulus square of the self-coherence function of the particle and the autocorrelation function of the detectors*. They are implicitely contained in the general formulae from (2.24) through (2.32) in a paper by Goldberger and Watson [17], for instance. Of course, one can denote any complicated expression by a single letter, as we have done here, too. But, on the other hand, one has to keep in mind that in the above derivation, we have been able to explicitely point to the part of the derivation where these complicated objects $M$ step on the scene. We shall present the explicit form of $M$ in very special limit forms, just for illustration purposes. These limit forms can also be derived from the correspondig formulas (6.1-20) and (6.1-21) in Chapter 6 of the book by Goodman [6], who also calls these quantities 'modes'. In case of a Gaussian transverse beam profile and a spatial detector of rectangular shape, we have approximately

$$M_G^x = \left[ \frac{1}{x} erf(\pi^{1/2} x) - \frac{1}{\pi}\left(\frac{1}{x}\right)^2 (1-e^{-\pi x^2}) \right]^{-1} \to \begin{cases} x, & x \gg 1 \\ 1, & x \ll 1 \end{cases}, \qquad x \equiv \frac{s_x}{\lambda_x} \times \delta_x, \qquad (22)$$

where $erf(z)$ denotes the error function, $\lambda_x$ is the transverse coherence length in the $x$-direction, and $s_x$ is the source dimension in the lateral (transverse) $x$-direction (perpendicular to the beam axis). A similar expression is valid for the spatial dependence in the $y$-direction, thus, the coherence area is defined as $\lambda_\perp^2 \equiv \lambda_x \lambda_y = A_c$. For gaussian temporal pulses, e.g. x-rays stemming from an undulator, the number of temporal modes is $\sqrt{1+(\tau_D/\tau_c)^2 \times \delta_t^2}$ [20, 21]. For electromagnetic radiation, due to the constancy of the velocity of light in vacuum we have $\lambda_{coh} = c\tau_c$. Expressed in simple approximate terms, the quantities $\delta_{x,y}$ measure the product of the number of elementary bundles of rays (degrees of freedom, in the sense used by von Laue) contained in the (solid) angles with centers at the two detector areas [6]. Of course, the Cartesian factorization $M = M_x M_y M_t$ is of limited validity, and must be applied with care. Here we use it only for illustating the essential features of the relevant modes in the HBT type counting experiments. In cases when the particles are created in an exponential decay process in the outer source and the detector time gate is rectangular, the overlap of the temporal autocorrelations can be approximated as

$$M_L = \left[ T^{-1} \int_{-\infty}^{\infty} \Lambda(\tau/T) |\gamma(\tau)|^2 \, d\tau \right]^{-1}, \quad \Lambda(x) = 0, \ |x|>1, \quad \Lambda(x) \equiv 1-|x|, \ |x|<1$$

$$= M_L = \left[ \frac{1}{x} - \frac{1}{2}\left(\frac{1}{x}\right)^2 (1-e^{-2x}) \right]^{-1} \to \begin{cases} x, & x >> 1 \\ 1, & x << 1 \end{cases}, \quad x \equiv \frac{2T_S}{\tau_c} \times |\delta_t|, \quad (23)$$

where $T_S$ is the effective duration of the gate (see formula (6.1-17) in Goodman's book [6]). In the experiments, the dimensionless 'overlaps' $\delta_x$, $\delta_y$ and $\delta_t$ are varied in the spatial and longitudinal directions when one searches for the boson 'bump' or the fermion 'dip' in the coincidence counts. In case of triggered counting, one is able to sweep e.g. the expected dinamics of a spontaneous (exponential) decay process, like radioactive decay or spontaneous light emission from a 'clean' source. In fact, this method can also be used for the measurement of the decay constants. The relative time overlap (delay) $\delta_t$ can both be negative or positive, but $\delta_{x,y}$ are to be concidered as non-negative *distances*. We emphasize, that the formulae in Equations (22) and (23) appear quite straightforwardly as multiple integrals in the usual formalism based on quantum or stochastic fields in calculating the fourth-order normally ordered moments of the type $\langle E^{(-)}(x_1)E^{(-)}(x_2)E^{(+)}(x_2)E^{(+)}(x_1) \rangle$, where $E^{(+)}$ and $E^{(-)}$ are the positive and negative frequency parts of the fields, respectively. Besides the calculation of the matrix elements of combinations of normally ordered products, the calculations always lead to the evaluation of spatio-temporal *overlaps of classical waves*. For example, in his book on photons and nonlinear optics, Klishko [54] introduces the 'important experimental parameter', the 'contrast parameter' $m$, which is the ratio of  true and accidental coincidences (the height of the 'bump' or the depth of the 'dip' in the coincidence curve). Under ideal circumstances this parameter is just the modulus squared of the self-coherence function, i.e. $m \approx |\gamma(\boldsymbol{r}_1, \boldsymbol{r}_2, 0, \Delta t)|^2$, where $\boldsymbol{r}_1$ and $\boldsymbol{r}_2$ are the positions of detectors, and $\Delta t$ is the applied delay. If the volume $V_{det}$ of the detection is considerably larger than the coherence volume $V_{coh}$ of the photon field, then $m = |\gamma(\boldsymbol{r}_1, \boldsymbol{r}_2, 0, \Delta t)|^2 \approx V_{coh}/V_{det} \equiv 1/g$. This means, that the parameter $g$ defined by Klishko [54], equals to our $M$, $g \approx M$. Thus, the 'contrast ratio' used by Klishko [54] can be considered as an approximate macroscopic average of our functional $M$. A more general mathematical definition can be obtained on the basis of the corresponding formulas derived by Goldberger, Lewis and Watson [14-18]. These formulas contain convolutions with the distributions of the detection points and the source points. In a mathematically strict sense the integral should be ment Lebesgue-Stiltjes integrals, defined by general Jordan-Lebesgue measures [39]. We note, in addition, that an analogouos quantity to our $M$ has formally appeared already in 1961, in the important study by Mandel and Wolf [5] on correlations in the fluctuating outputs from two square-low detectors. Recently, in the context of counting experiments, the name 'mode' has also been systematically used by Yabashi *et al.* [19-20] and Ikonen *et al.* [21] in interpreting the results of their recent experiments on intensity-intensity correlations of x-rays coming from an undulator. The limit expressions of $M_G(x)$ and $M_L(x)$, and, in fact, the functions themselves can be quite well approximated by the function $1+x$, as is shown in Figures 2 and 3.

SPACE FOR FIGURE 2

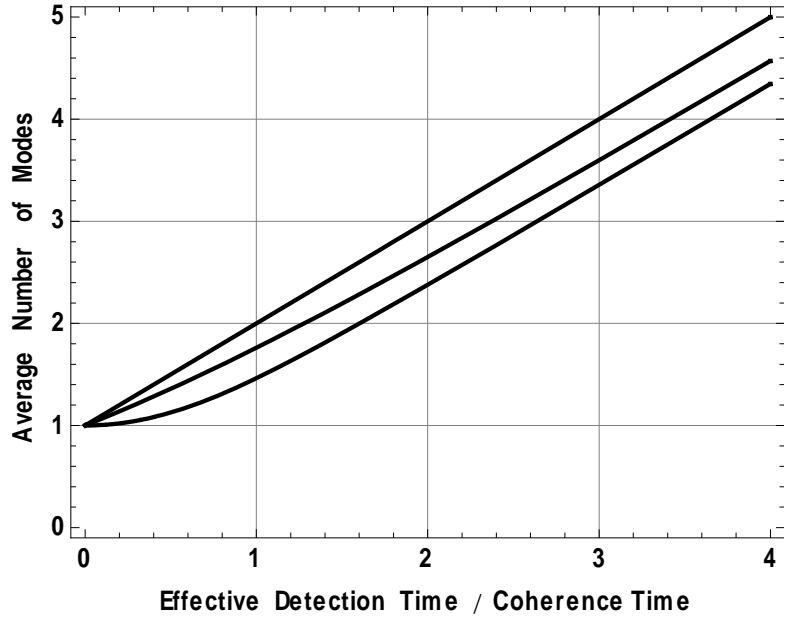

**Figure 2.** Shows a comparison of the number relevant modes as a function of the relative overlap in time delay experiments. On the abscissa the fractional overlap $x \equiv (T/\tau_c) \times |\delta_t|$ is varied by changing the temporal occupation number of quanta from one sequence to the other. On the ordinate the function values of $M_L(x)$ given by Equation (23) are displayed by the middle curve, and the lower curve refers to the Gaussian function $M_G(x)$, given by Equation (22). Here the varied parameter is $x \equiv (s_x / \lambda_x) \times \delta_x$, which quantifies the spatial fraction along the $x$-direction which the quantum covers within the lateral spacing of the detectors. For a comparison, we have plotted the straight line corresponding to the function $1 + x$, which is the uppermost line. According to Equations (22) and (23) the curves are practically coincide with $x$. On the figure we have plotted $1 + x$, rather, because we wished to show that a quite simple approximation can also be used if the mode number is not so much larger than 1.

SPACE FOR FIGURE 3

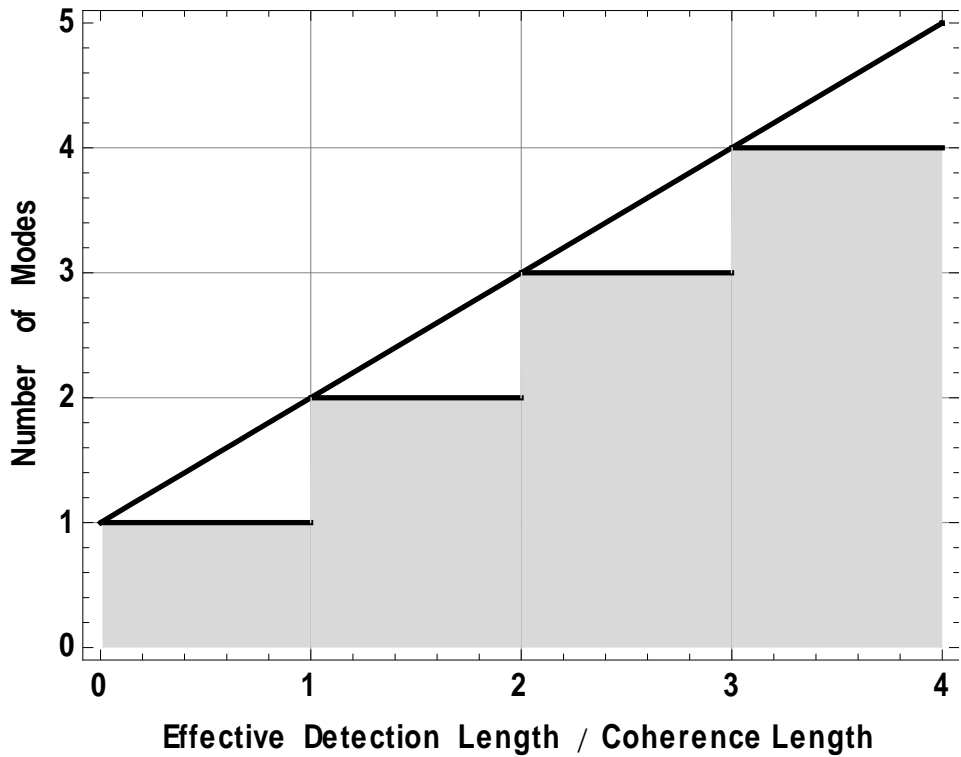

**Figure 3.** Shows the function $1+x$ and its integer part, $[1+x]$, where $x \equiv (s_x/\lambda_x) \times \delta_x$ is a lateral fraction in the $x$-direction perpendicular to the beam axis. Though, in a strict sense, the number of degrees of freedom, $M$ is a positive integer, its value cannot be controlled sharply in real experiments. The unsharpness of the number of modes may mean, for instance, the unsharpness of the geometrical boundary conditions. This may be considered as a source of decoherence.

In the light of the above considerations, we can state with certainty, that in many cases of linear two-point experiments the measured effect is a self-coherence effect, since otherwise the observation of true coincidences could not have been explained, due to the low degeneracy of the beam. The true (anti)coincidences can be associated to the detection of single quanta by two shifted copies of the detectors. The correlation between the detection events stems from the fact that in a highly monoenergetic dilute beam all the single (and well separated) quanta have the same coherence length (volume). Each copies of the detectors absorb only one quantum at a time, but the statistics of these absorption processes are connected by the overlapping of the coherence volumes (of exactly the same sizes) with the intersection of the two shifted copies. At zero mismatch the whole interaction region covers completely the coherence volume of one quantum, and then the contrast of the coincidence curve can reach its maximum value, namely unity. This is the case when the number of 'relevant modes' is just the minimum, namely 1. The phenomenon is illustrated in Figure 4 for bosons and fermions.

SPACE FOR FIGURE 4

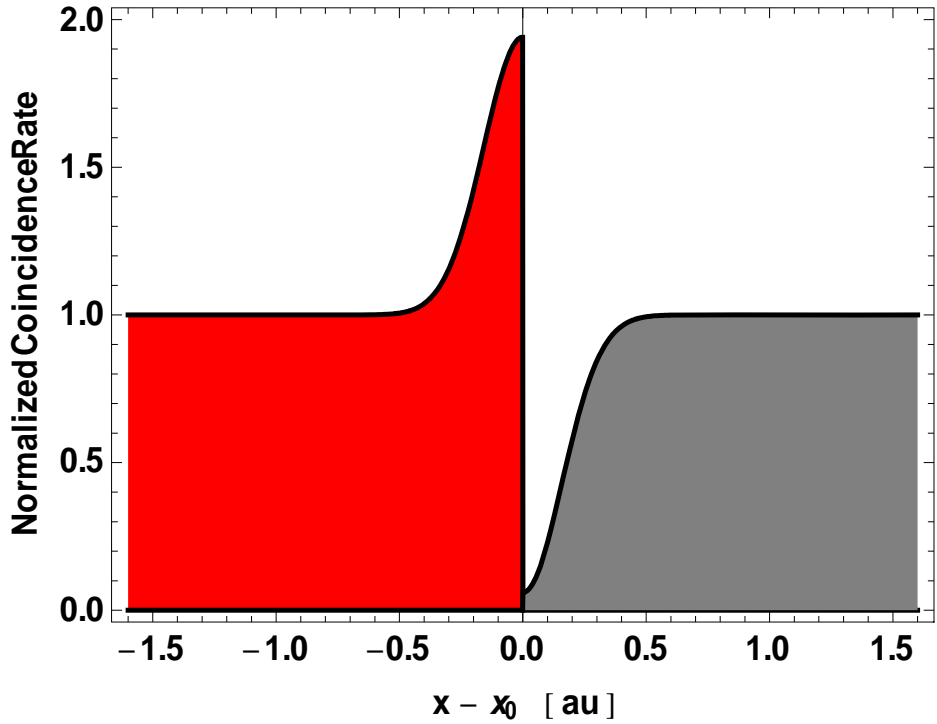

**Figure 4.** Shows ideal two point bunching and antibunching curves illustrating the normalized coincidence rates at two detectors for bosons (left curve with red filling) and fermions (right curve with gray filling) as functions of the spatial mismatch $|x - x_0|$ between the two detectors, according to Equations (18) and (20). The curves illustrate the appearance of excess (anti)coincidences in comparison to the poissonian background. In making this figure, a Gauss approximate of the profile function in Equation (22) has been used. The height of the bunching 'bump' and the depth of the fermion 'dip' is 1, and their width is also 1 in this ideal case. At zero mismatch the whole interaction region covers completely the coherence volume of one quantum, and then the contrast of the coincidence curves can reach their maximum value, namely unity. This is the case when the number of 'relevant modes' is just the minimum, namely 1. The rate of accidental (classical) coincidences equals to 1, corresponding to the 'very large number of modes limit' characterized by the Poisson weights in Equation (21).

## 5. The analysis of two recent experiments on HBT type correlations of massive particles, and further general considerations on the relevant modes

In order to illustrate our usage of the concept of modes (in the completely general sense, as degrees of freedom) in the context of two-point correlation experiments, we discuss here two examples. In a recent important experiment, Jeltes *et al.* [25] have made a direct comparison of HBT type correlations in fermionic $^3$He beams and bosonic $^4$He beams by releasing a cold cloud of metastable helium atoms at the switch-off of a magnetic trap. They observed ~6% relative decrease and ~3% increase of the counts for the fermionic and for the bosonic components, respectively. They interpreted the difference in the sizes of the fermion 'dip' and of the boson 'bump' by the different longitudinal coherence lengths of the isotopes. They estimated that the $^3$He fermions have a larger coherence length ~0.75mm and the $^4$He have ~0.56mm, and they explained the difference between the contrasts by the difference between the coherence lengths. According to the measured numbers, the difference factor (suspectedly very close to 2) cannot come out merely on the basis of such an argumentation. We have made a numerical comparison, given by our general formulas in Equations (18) and (20), with the experimental result of Jeltes *et al.* [25], and found first a quatitatively good agreement by calculating the number of lateral modes. Of course, in the neighborhood of the dip or bump they swept one longitudinal mode, whose width is naturally governed by the longitudinal coherence length. As for the difference in the contrast depths for bosons and fermions, we think that the factor of $2 = (6\%)/(3\%)$ may simply show up as a consequence of the preparation of the particles in the magnetic trap. At the moment of the switching–off of the trap the fermions were polarized, and the coincidence dip was 2-times deeper for the polarized fermion than for the evidently unpolarized bosons. Thus, the difference is the consequence of the reduction of the number of modes (degrees of freedom) by a factor of 2, and this may explain the experimental result.

The experimental results of Iannuzzi *et al.* [31] on antibunching, observed with thermal neutron beams, can also be quantitatively interpreted on the basis of our general formula in Equation (20) for fermions. In a HBT type experiment, in which one uses a beam–splitter and shifts one of the detectors along the direction of the impinging beam, this shifting is not equivalent with the detection of delayed coincidences along the beam axis. This is because the different components of the momenum are connected by the energy expression (dispersion relation). If the source is chaotic laterally, then the shape of the bunching 'bump' (or 'dip') is essentially Gaussian in both cases. However in the first case the center of the bump should be calibrated to zero, simply because in reality there exist no negative distance. If one uses both negative and positive spatial shifts, then the coordinate of the calibration point must be positive, and possibly larger than the width of the bump or dip, in order to see a measurable effect. The importance of this point has also been expressed by Iannuzzi *et al*. [31]. Concerning this experiment, there are two points to be mentioned briefly here. The possibility of measuring neutron antibunching has first been discussed by Boffi and Cagliotti [26] in 1966 and later in 1971 (see Refs. [26] and [34]), on the basis of the general analysis of Goldberger, Lewis and Watson [14-18], but the feasibility of such an experiment has been seriously questioned for a long time. The reason is that, Silverman [26] in one of his important papers on fermion correlations (antibunching), published in 1988, claimed to had proved that measuring neutron antibunching, with presently available neutron sources, would practically be impossible. The reason for that would be the extreme smallness of the degeneracy parameter of thermal neutron beams. Really, the average occupancy of neutrons is around $10^{-14}$ in one phase-space cell (coherence volume × spectral volume), even in the nowadays most intense thermal sources [34]. In case of such a low degeneracy one does not expect that the *two-particle fermionic character* can manifest itself, as for instance statistical repulsion. According to Silverman [26], in order to have an acceptable signal to noise ratio (which is proportional to the degeneracy parameter), one would need 10000 years, or even more, for a reliable complete experiment. His statement is correctly supported by the mathematical analysis, but he has not taken into account, that the beam splitter may play the role of the source in the scattering experiment, under certain conditions. This possibility had already been discussed by Goldberger and Watson long ago [17]. In fact, in the experiment [31], the beam splitter was an opaque object for the impinging neutrons, which were mosaic crystals with 'grains' of the size on the order of ~1μ$m$. Now, according to the Van Cittert-Zernicke theorem [6], one should take into account that the area of the effective source (the mosaic elements as elastic scatterers) covered only few tens of coherence areas, thus, the effective degeneracy may have been of the order of $10^{-5} - 10^{-4}$, according to our estimates. The true *coherent* two-particle effects still cannot, but single-quantum self-correlations can already be observed with reasonable data acquisition times. In the experiment the drastic supression of the number of relevant modes considerably reduced the data acquisition time, by increasing the signal to noise ratio to an acceptable level. By applying the second equation of Equation (20), with $M = M_s M_t$, $M_t \approx \tau_D / \tau_c$, $M_s \cong A_s / A_c \sim 26 - 30 \sim (2) \times 14$, we have a reasonably good agreement with the experimental results. Here $\tau_D$ and $\tau_c (= \lambda_c / \upsilon)$ are the effective time window (including the dead time of the detectors) and the coherence time of the neutrons, respectively. $\lambda_c$ is the associated coherence length, and $\upsilon$ is the central velocity of the neutrons. $A_s$ denote target area (i.e. the effective source in the beamsplitter) and $A_c$ is the coherence area of the neutron beam. This simple approximate formulae can be derived under the assumption of cross-spectral purity, and by using the expression in Equation (22). The requirement of cross-spectral purity has certainly been satisfied in this experiment, because the neutron beam was quite monochromatic ($\Delta E / E < 10^{-3}$), thus, the self-dispersion was negligible during the propagation. In other words, there was no considerable 'cross-talk' of the spatial and temporal evolution. Besides, we think that the mosaic crystal grains in the graphite beam splitter did not bring in considerable dispersion, because they played the role of elastic scatterers, which redistributed the energy and phases merely among the spatial modes. Our numerical estimates essentially coincides with that of the authors, but the physiscal interpretations are completely different. For instance, in our formalism, it naturally comes out that the depth and the witdth of the coincidence curves are in general, two independent parameters. The former is related to the number of lateral modes, if we probe the beam longitudinally, by shifting one of the detectors. In one of the experiments of Iannuzzi et al. [31], these quantities *accidentally* were the same. Concerning the role of the mosaic crystals, it is interesting to note that in the original experiments of Hanbury Brown and Twiss, the starlight in the stellar intensity interferometer was focused to the detectors by parabola antennas whose reflecting surfaces consisted of mosaics of plane mirrors. In this way, on one hand, the effective flux of the collected starlight was increased at the detectors, and, on the other hand, the number of relevant spatial modes was reduced, yielding a considerable increase of the contrast of the signal, according to our Equation (18). It is also interesting to compare the contrasts of the neutron antibunching curves measured by Iannuzzi *et al*. [31] by using two different kinds of detectors, namely the $^3$He gas detectors and the scintillators. According to our general formula given by Equation (20), the relative size of the antibunching dip can be quantified by the ratio $N_{bg}/(N_{bg} - N_c) = Z$, where $N_{bg}$ and $N_c$ are the number of background (or 'accidental') and 'effect-coincidences', respectively, and $Z$ is either $M$ or $2M$, depending on the polatization degree of the detected particle beams. From the quoted work [31] we have calculated $N_{bg} \cong 34720$ and $N_c \cong 34480$ at the centre of the dip, measured with the $^3$*He* gas detectors, and $N_{bg} \cong 994$ and $N_c \cong 960$, measured with the scintillator detectors, respectively, yielding

$$Z \equiv \frac{N_{bg}}{N_{bg} - N_c} \approx \begin{cases} Z_1 = 14, & {}^3He \ gas \\ 2 \times Z_1 = 28, & scintillator \end{cases}. \quad (24)$$

Equation (24) shows that the antibunching dip is shallower by a factor of 2 with scintillator measurements (i.e. from the background we subtract only $1/(2Z_1)$ intead of $1/Z_1$). According to our description, the number 14 is just the number of coherence areas of the radiation, stemming from a mosaic region, covered by the detectors. The difference of the factor of 2 is due to the use of different detector materials (which, of course, affects the absorption lengths, too, as was mentioned by the authors). We suspect that, since the absorber ${}^3He$ nuclei are fermions, in the coincidence region mostly (incoherent!) neutron pairs with the same polarization are counted (taken into account in the number of coincidences). According this rather sketchy argumentation and to our formulas in Equation (20), for the components with the same polarization we have to have a 2 times larger contrast, so this can be one simple explanation of the very different contrasts found by using the two different detectors.

At the end of the present section we would like to make still two remarks on the physical nature of the 'relevant modes'. In Equations (22) and (23) we have given mathematical expressions, which naturally come out in the quantum mechanical analysis of two-point correlation experiments. We stated that these are in fact just the number of relevant modes $M$, which we have used in the algebraic treatment in the framework of classical probability theory. With this association, our general results coincide with that of usual quantum mechanical calculations. We are well aware of that, that this procedure, at this level of the presentation, could simply be considered as mere an *a posteriori* justification. In the present paper we do not aim to give the mathematically precise study of the possible equivalence. We plan to deal with these details in a separate publication [39]. We note that Kelley and Kleiner [75] have given a thorough analysis of electromagnetic field measurements and photoelectron counting in terms of compound Poisson distributions [46] relying on the P-representation of the quantized fields. Our present method is different to theirs, e.g. also in that respect that the basic algebra of the elementary microscopic sequences is described by trinomial distributions, rather than elementary binomial stochastic processes. As we have seen above, the measured relative frequencies quantify the *fluxes of counts*, whose sum is not constant, just because of the partial reflection at the entrance port. According to the present description, the freedom received by the use of the trinomial distribution makes it possible to deal with the elementary correlations in more details, and keep track of the algebra of events, which is particularly important in the single-photon regime. This is not possible on the basis of the strict binomial dichotomy.

Finally we summarize some simple mathematical expressions, on the basis of which the physical meaning of $M$ can be understood. The magnitude of the functions $M$ in a given experiment can be well estimated according to the following simple calculations. Take, for example nonrelativistic particles of energy $E = p^2/2m$ and velocity $\upsilon = p/m$ covering the length $L = \upsilon \tau_D = M \Delta x$, which is divided into $M$ pieces. Writing $\Delta x = h/\Delta p$, we have $M = \tau_D p \Delta p / mh$, thus we obtain $M = \tau_D \Delta E / h$, i.e. $M = \tau_D \Delta \nu = \tau_D / \tau_c$, which is just the value of the modes on the right hand side of Equation (23). A similar argument leads to the same result if we consider photons, and take into account the dispersion relation $p = \varepsilon/c = h\nu/c$. Thus, in general, we can write for both photons and massive particles the asymptotic expression (for large $M$) for the number of relevant longitudinal modes:

$$M \approx \tau_D \Delta E / h = \tau_D \Delta \nu = \tau_D / \tau_c. \tag{25}$$

After Oxborrow and Sinclair [82], Scheel [85] has formally defined the "*mutual indistinguishability*" $\widetilde{M} = 1/M$ in terms of the combination of second order quantum coherence functions, which is exactly the inverse of our $M$. Thus that quantity, at the same time, is a variant of Klishko's [54] (1/ 'contrast ratio'). According to Scheel [85], "By definition, the depth of the Hong-Ou-Mandel dip is proportional to $1 - \widetilde{M}$."

The counting events are local, not only in a practical but in a strict sense, and we can attach to each detection point an associated coherence region. The self-coherence of each single quantum cause the correlations of detections at two different spatio-temporal points, because the quanta cannot be distiquished within a coherence volume. The number of relevant modes has the clear physical meaning of being the relative intersections of ratios of the interaction volume and the coherence volume of the quantum. As we shall see in the following section, in single-photon beam splitter experiments, this intersection is the product of the relative number of particles counted by detector A and B on the opposite sides of the beam splitter, multiplied by the average normalized flux of the pumping of the coincidence apparatus.

## 6. Application of the general results for the interpretation of photon anticorrelation effects in 'single-photon experiments'

In the photon correlation experiments performed by contolling very low-intensity photon beams (in the single-photon regime by using single-photon sources) 'photon anticorrelation effects' show up, as has first been demonstrated by Kimble, Dagenais and Mandel [51] and Aspect, Grangier and Roger [58] in their fundamental experiments (see also e.g. the recent works by Jaques *et al*. [59], and [79-85]). In our terminology, this kind of expriments belong to the special cathegory of few sequence whole experiments, where the average photon occupation number can be considerably less than 1, and, at the same time, the average total number of modes is also around unity *at the minimum* of the anticoincidence curve. The contrast ratio (visibility or modulation) of the coincidence curve is always on the order of $1/M_{min}$, in both the standard theory and in the present one (see the discussion above on the connection of the 'contrast ratio' $m = 1/g$ introduced by Klishko [54] and the quantity $M \approx g$). Under the condition of cross-spectral purity, the mode number is given as $M = M_{tr}M_{l}$, where $M_{tr}$ and $M_{l}$ are the transverse and longitudinal mode numbers, respectively. For instance, in the case when only one transverse mode is sampled, and the resolution is high enough to resolve one longitudinal mode, then the contrast ratio reaches its maximum, namely 1.

As the last example in the present paper, we shall now analyse the experiment by Aspect and Grangier [58], on the basis of results obtained by classical probability theory in the preceding sections. After summarizing the basic points to be kept in eye in this context, we shall attemp to show that the non-classical nature of light does not prevent us from a quantitatively accurate classical description of the anticorrelation phenomena, without the use of second-quantized field amplitudes. We would like to point out again, that we are not attempting to replace the quantized amplitudes with infinitely divisible classical stochastic fields, but we describe the discrete counting events by random integer variables defined on the Boole algebra associated to a particular experiment.

In order to have a brief overview of the theoretical and experimental aspects concerning such phenomena, we think the best is just quoting the relevant parts in subsection 4b of Ref. [58]; "An excited atom emits a single photon, because of energy conservation. In classical sources, many atoms are simultaneously in view of the detectors, and the number of atoms fluctuates. As a consequence, the emitted light is described by a density matrix reflecting these fluctuations, including the possibility that several photons are emitted simultaneously. For a Poisson fluctuation of the number of emitting atoms, one can show that the statistical properties of the light cannot be distinguished from the one of classical light. In order to observe non-classical properties in fluorescence light, it is thus necessary to isolate single atom emission. This was realized by Kimble et al. [[51]] who had only one atom in their observation region when they demonstrated antibunching. In our experiment, we have been able to isolate single atom emission not in space but in time. Our source is composed of atoms that we excite to the upper level of a two-photon radiative cascade (fig. 3) [[58]], emitting two photons at different frequencies $\nu_1$ and $\nu_2$. The time intervals between the detections of $\nu_1$ and $\nu_2$ are distributed according to an exponential law, corresponding to the decay of the intermediate state with a life time $\tau_s = 4.7 ns$. By choosing the rate of excitation of the cascades much smaller than $(\tau_s)^{-1}$, we have cascades well separated in time. We can use the detection of $\nu_1$ as a trigger for a gate of duration $w \cong 2\tau_s$, corresponding to the scheme of fig. 2. During a gate, the probability for detection of a photon $\nu_2$ coming from the same atom that emitted $\nu_1$ is much bigger than the probability of detecting a photon $\nu_2$ coming from any other atom in the source. We are then in a situation close to an ideal single-photon pulse [11], and we expect the corresponding anticorrelation behaviour on the beam splitter." In the photon anticorrelation experiments performed by Aspect, Grangier and Roger [58] the average photon occupation number was very low due to the low incoming energy flux. As they wrote: "The expected values of the counting rates can be obtained by a straight-forward quantum mechanical calculation. Denoting $N$ the rate of excitation of the cascades, and $\varepsilon_1$, $\varepsilon_t$ and $\varepsilon_r$ the detection efficiencies of photon $\nu_1$ and $\nu_2$ (including the collection solid angles, optics transmission, and detection efficiencies) we obtain

$$N_1 = \varepsilon_1 N \tag{7a}$$

$$N_t = N_1 \varepsilon_t [f(w) + Nw] \tag{7b}$$

$$N_r = N_1 \varepsilon_r [f(w) + Nw] \tag{7b'}$$

$$N_c = N_1 \varepsilon_t \varepsilon_r [2f(w)Nw + (Nw)^2]. \tag{7c}$$

The quantity $f(w)$, very close to 1 in this experiment, is the product of the factor $[1 - \exp(-w/\tau_s)]$ (overlap between the gate and the exponential decay) by a factor somewhat greater than 1 related to the angular correlation between $\nu_1$ and $\nu_2$ [12]. The quantum mechanical prediction for $\alpha$ (eq. (6)) is thus

$$\alpha_{QM} = \frac{2f(w)Nw + (Nw)^2}{[f(w) + Nw]^2} \tag{8}$$

which is smaller than one, as expected. The anticorrelation effect will be stronger ($\alpha$ small compared to 1) if $Nw$ can be chosen much smaller than $f(w)$. This condition corresponds actually to the intuitive requirement that $N$ is smaller than $(\tau_s)^{-1}$ ($w$ is of the order of $\tau_s$). The counting electronics, including the gating system, was a critical part in this experiment. The gate $w$ was actually realized by time-to-amplitude converters followed by threshold circuits. These single-channel analyzers are fed by shaped pulses from PM1 (detecting $\nu_1$) on the START input, and from PM$_r$ or PM$_t$ on the STOP input. This allows us to adjust the gates with an accuracy of 0.1ns. A third time-to-amplitude converter measures the delay between the various detections, and allows to build the various time delay spectra, useful for the control of the system." The reference [12] quoted by Aspect and Grangier is the paper by E. S. Fry: Two-photon correlations in atomic transitions. *Phys. Rev. A* **1973**, *8*, 1292-1232. The parameter $\alpha$ had been defined in section 3 of Ref. [58] in the following context: "Therefore, for any classical-wave description of the experiment of fig. 2, we expect

$$P_c \geq P_r P_t \tag{6a}$$

or equivalently

$$\alpha \geq 1 \quad \text{with} \quad \alpha = \frac{N_c N_1}{N_r N_t}. \tag{6b}$$

The intuitive meaning of this inequivality is clear. For a classical wave divided on the beam splitter, there is a minimum rate of coincidences, corresponding to the 'accidental coincidences'. We have thus obtained a criterium for empirically characterizing a single particle behaviour of light pulses. The violation of inequality (6) will indicate that the light pulses should not be described as wave packets divided on a beam splitter but rather as single photons that cannot be detected simultaneously on both sides of the beam splitter." In eq. (6) of Ref. [58] $P_t$, $P_r$, $P_c$ and $N_t$, $N_r$, $N_c$ are the transmitted, reflected and coincidence probabilities and fluxes, respectively, and $N_1$ denotes the trigger rate.

On the basis of classical probability theory of radioactive cascades [45], it is a simple matter to obtain the population probability of the intermediate level (level (2)) of the cascade transition

$$P_2(t) = \frac{\gamma_1}{\gamma_2 - \gamma_1}(e^{-\gamma_1 t} - e^{-\gamma_2 t}), \tag{26a}$$

where $\gamma_1$ and $\gamma_2 = (\tau_s)^{-1} = (4.7 ns)^{-1}$ are the decay constants of the upper and of the intermediate level, respectively. In the experiment the condition $(\tau_s)^{-1} = \gamma_2 \gg \gamma_1$ has been secured, in order to separate the two transitions in the cascades of the *same* atoms. The mathematical expression for $P_2(t)$ in Equation (26) has two physical meanings in the experiment under discussion. On one hand, being the probability that one atom is in the intermediate level (in other words, 'after the first transition (1) → (2) in the cascade has already been taken place'), $P_2(t)$ serves as a proportionality factor in the probability of observing the trigger photon $\nu_1$ within a solid angle $\Omega_1$,

$$\mathrm{Pr}_{obs}(\nu_1;t) = \eta_1 \Omega_1 P_2(t), \qquad P_2(t) \approx \frac{\gamma_1}{\gamma_2}(1 - e^{-t/\tau_s}), \qquad \mathrm{Pr}'_{pump}(\nu_2;t) = \eta_{pump} \Omega_2 P_2(t), \tag{26b}$$

as is shown in the first equation of Equation (26b). On the other hand, $P_2(t)$ is approximately proportional to the emission probability of a photon $\nu_2$ stemming from the second transition (2) → (3) of the same cascade. For $t \ll \tau_s$ we have $P_2(w) \approx \gamma_1 w$ during a gate, and this gives a pumping term $Nw$, where $N$ denotes the pump rate of the upper level of the cascades. The third equation represents the pumping term coming from the isolated spontaneous emission process (2) → (3) during a gate, which would start at $t = 0$. Thus, in fact, the pumping can be represented as a sum of two terms $Nw + f(w)$, where, according to the third equation of Equation (26b), $\tilde{f}(w) = (\Omega_2/\Omega_1)(1 - e^{-w/\tau_s})$, and $\Omega_2$ is the solid angle within which the excited sample of the cascade atoms (sending the photons $\nu_2$) are seen from the entrance port of the coincidence apparatus. $\Omega_1$ has been defined in Equation (26b). The function $\tilde{f}(w)$ should essentially be the same function as $f(w)$ in eqs.(7) of Ref. [58]. In the experiment the duration of the electronic gates were optimized so that the value numerical value of $f(w)$ was kept close to unity, i.e. $e^{-w/\tau_s} \ll 1$. If we plot the function (26a) for $\gamma_1 \approx 0.1\gamma_2$ in the interval (0,1) with $w/\tau \approx 2$, we see that it is essentially a linear function of its dimensionless argument.

Since, on the average, definitely less than <1/10 photon is in the apparatus during a gate, the statistical weights in the general formula given by Equations (8) and (9) can be taken as

$$W_n(M) = \binom{M}{n} (\bar{n})^n (1-\bar{n})^{M-n}, \quad (0 < \bar{n} \leq 1), \tag{27}$$

where $\bar{n}$ is the mean occupation number. In Equation (27) we have a binomial distribution, with which we have already encountered in Equation (19a). According to the general formula given by the first equation in Equation (20), the normalized counts are obtained,

$$K = \frac{\overline{\xi \cdot \eta}}{\bar{\xi} \cdot \bar{\eta}} = 1 - \frac{1}{M}, \quad M = \left[\frac{1}{x} - \frac{1}{2}\left(\frac{1}{x}\right)^2 (1 - e^{-2x})\right]^{-1}, \quad x \equiv \frac{2T}{\tau_c} \times \delta = \frac{4T}{\tau_s} \times \delta, \tag{28}$$

where the Lorentzian mode function given by Equation (23) has also been used. According to the arguments followed by Equation (26b), for the dimensionless overlap we have $4(T/\tau_s)\delta = 4 \times Nw \times \tilde{f}(w)$. Here $Nw$ is the average number of cascades during a gate of duration $w = T$, and $\tilde{f}(w) = (\Omega_2/\Omega_1)(1 - e^{-w/\tau_s})$ has already derived above. The factor $1 - e^{-w/\tau_s}$ is the probability that an emission of a photon $\nu_1$ from the intermediate level of the cascade takes place during $w$, and in the first factor $\Omega_2$ and $\Omega_1$ are the solid angles of the collection of the trigger photons and that of the measured photons, respectively. The data acquisition has been performed for seven values of the average number of cascades during a gate. In our numerical calculations we have used the values of $Nw$ = {0.06, 0.12, 0.18, 0.3, 0.54, 0.75, 1}, taken from the values of the abcissas of the measurement points in Figure 4 of Ref. [58], and we have taken $w/\tau_s = 9/4.7$ and $\Omega_2/\Omega_1 = 1.06$ in accord with the experiment. With these parameters we have $\tilde{f} = 0.9$ and $4(T/\tau_s)\delta = 3.6 \times Nw$. The numerical result obtained on the basis of Equation (28) are summarized in Table 1. The numbers in the sixth column have been derived from the quantum formula for $\alpha_{QM}$ given by eq. (8) in Aspect and Grangier [58], for which the numerical result is $\alpha_{QM}$ = { 0.1211, 0.2215, 0.3056, 0.4375, 0.6094, 0.7025, 0.7756 }. The number of coincidences calculated on the basis of our formula for $K$ in Equation (28) is shown in the seventh coulumn, where $K$ = { 0.1297, 0.2352, 0.3217, 0.4533, 0.6152, 0.6979, 0.7608 }. The relative differences $100(\alpha_{QM} - K)/(\alpha_{QM} + K)$ of these results is about { -3%, -3%, -3%, -2%, 0.5%, 0.3%, 1% }, respectively.

SPACE FOR TABLE 1.

| Reflected singles $n_{2r}$ | Transmitted singles $n_{2t}$ | Expected(1) coincidences | Expected(2) coincidences | Measured coincidences | Calculated(1) coincidences | Calculated(2) coincidences |
|---|---|---|---|---|---|---|
| 2940 | 3876 | 25.5# | 2* | 6 | 3# (0.24)* | 3# (0.26)* |
| 78260 | 95840 | 50.8 | 49 | 9 | 11 | 12 |
| 91908 | 124912 | 64.1 | 64 | 23 | 20 | 21 |
| 241920 | 326400 | 204 | 202 | 86 | 88 | 91 |
| 409200 | 535920 | 456 | 455 | 273 | 277 | 279 |
| 399840 | 519960 | 492 | 492 | 314 | 346 | 343 |
| 257400 | 344880 | 367 | 367 | 291 | 285 | 280 |

**Table 1.** Gives a comparison of the experimental data of Aspect and Grangier [58] on single-photon anti-correlation with the theoretical results quoted by the authors and with that of the present work. The seven raws correspond to the total number of coincidence gates $n_g = N_1 T$ =$10^3 \times${5664, 152564, 179080, 391680, 481800, 422520, 241560} during which the number of counts were registered. We have calculated these numbers on the basis of the experimental data given by the authors, namely, we

have taken for the trigger rates $N_1$ = {4720, 8870, 12100, 20400, 36500, 50300, 67100}sec$^{-1}$, and for the gate durations $T$ = {1200, 17200, 14800, 19200, 13200, 8400, 3600}sec, as have been given by the authors in the first and fourth columns in Table 3 in their paper. In the present table, in the first and the second columns, the calculated number of reflected photons, $n_{2r} \equiv N_{2r}T$, and the calculated number of transmitted photons, $n_{2t} \equiv N_{2t}T$, are shown, respectively. The numerical values of the fluxes $N_1$, $N_{2r}$, $N_{2t}$ and the durations of the data acquisition $T$ have been taken from Table 3 of the original reference [58]. In the third column (with heading "Expected(1) coincidences") the calculated number of accidental coincidences $N_{2r}N_{2t}T/N_1$ are shown, as has been given in Ref [58]. These would be the number of joint counts, which is expected according a classical Poisson background of rare events. In the fourth coloumn (with heading "Expected(2) coincidences") we show the number of accidental coincidences calculated by us. These numbers should exactly be identical with $N_{2r}N_{2t}T/N_1$ given in Ref. [58]. By going over to particle numbers, the duration $T$ drops out, and we have $n_{rt}^{acc} = (n_{2r}) \cdot (n_{2t})/(n_g)$ (we have displayed the integer parts). These numbers are expected in the measurement, according to the usual assumption of a Poissonian background of rare events, which may result only in "accidental coincidences". Except for the values 25.5$^\sharp$ and 2* in the first raw, the expected numbers given by the authors of Ref. [58] and calculated by us, are the same. We have denoted the value 25.5 displayed in Table 3 of Ref [58] by the symbol $\sharp$, because we think that it is due to a mistype or calculation error, because the general formula used by the authors of Ref. [58] and by us gives $n_{rt}^{acc} = 2$ in this case. In the sixth column (with heading "Calculated(1) coincidences") we list the number of relative coincidences we have calculated on the basis of the formula given by Equation (28), and presented by the authors [58] (see their equation (8) quoted above). These theoretical values have not been presented by Aspect and Grangier [58], but for us they served as an important comparison. In the seventh column with heading "Calculated (2) coincidences" we presented our results on the basis of Equation (28).

As is well-known, the experimental data clearly proved the anti-correlation effect. In Fig. 4 of Ref. [58] the experimental results are shown, and the $\alpha_{QM}$ – curve is also displayed, which follows quite accurately the measured points. "The indicated error is $\pm 1$ standard deviation.", according to the figure caption. "The value of $\alpha$ is $0.18 \pm 0.06$, corresponding to a total number of coincidences of 9, instead of a minimum value 50 expected for a classical model of light." ] In the fifth column the calculated number of coincidences are shown, according to the quantum field theoretical formula given by Aspect and Grangier [58]. The numbers in the 6$^{th}$ and in the 7$^{th}$ columns of Table 1. have been calculated by muliplying the "Expected coincidences", i.e. $n_{rt}^{acc} = (n_{2r}) \cdot (n_{2t})/(n_g)$, with $\alpha_{QM}$ and $K$, respectively. The theoretical results in the seventh column have been obtained on the basis of our formula given by Equation (28). In the first raw of Table 1 we see that for very low pump rate the measured coincidence number is 3 times *larger* than the background, thus this point corresponds to positive correlation, rather than to negative correlation. Such a transition from anticorrelation to bunching has recently been observed by Hennrich, Kuhn and Rempe [83]. According to Eqs. (22-26) and the 3$^{rd}$ raw and 4$^{th}$ column in Table 1 in our earlier paper [36], the factor of 3 here corresponds to the value $\varepsilon a^2 = (\sqrt{2}-1)/2$, where $\varepsilon \ll 1$ is the squeezing parameter and $pa^2$ corresponds to the average number of counts in one arm of the beam splitter.

The supression of coincidences characterized by the normalized counts of coincidences (the parameter $\alpha_{QM}$ of Ref. [58], see eq. (8) in the quotation above) is displayed on Figure 5 as a function of the overlap of the detection time and the spontaneous decay process. This overlap was changed by increasing the trigger rate through increasing the average population of upper level of the source atoms undergoing a two-step cascade decay.

SPACE FOR FIGURE 5

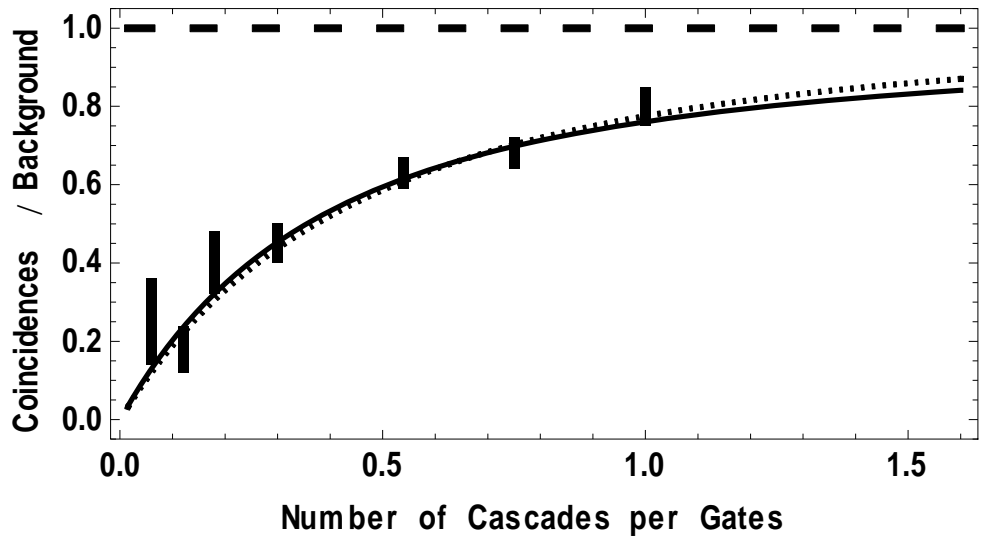

**Figure 5.** Shows the photon anticorrelation curve calculated by us, according to Equation (28). The value of $K = 1 - 1/M$ has been plotted, by using the experimental parameter $w/\tau_s \approx 9ns/4.7ns$. On the horizontal axis the dimensionless pump parameter $Nw$ is varied from zero to the value 1.6. The figure has been drawn without using any fitting. The dotted line shows the theoretical value of $\alpha_{QM}$, and the rectangles (error bars) represent the experimental results of Ref. [58]. The coordinates of the experimental results have been taken from Fig. 4 of Ref. [58]. In this figure the upper and lower ends of the error bars have the coordinates {{0.06, 0.25±0.11}, {0.12, 0.18±0.06}, {0.18, 0.40±0.08}, {0.3, 0.45±0.05}, {0.54, 0.63±0.04}, {0.75, 0.68±0.04}, {1., 0.80±0.05}}. The agreement between $\alpha_{QM}$ and

$K$ is almost perfect (as can also be seen in the sixth and the seventh columns in Table 1.), moreover, both theoretical values reproduce quite well the experimental results.

The coincidence curve calculated by us, and shown in Figure 5 practically coincides with the curve calculated by the quantum formula. Within the error bars each formula give back quantitatively the experimental results. In Figure 6 we show the dependence of the number of relevant modes in the experiment, as a function of the dimensionless parameter $Nw$. In Ref. [58], the maximum value of this parameter was 1.

SPACE FOR FIGURE 6

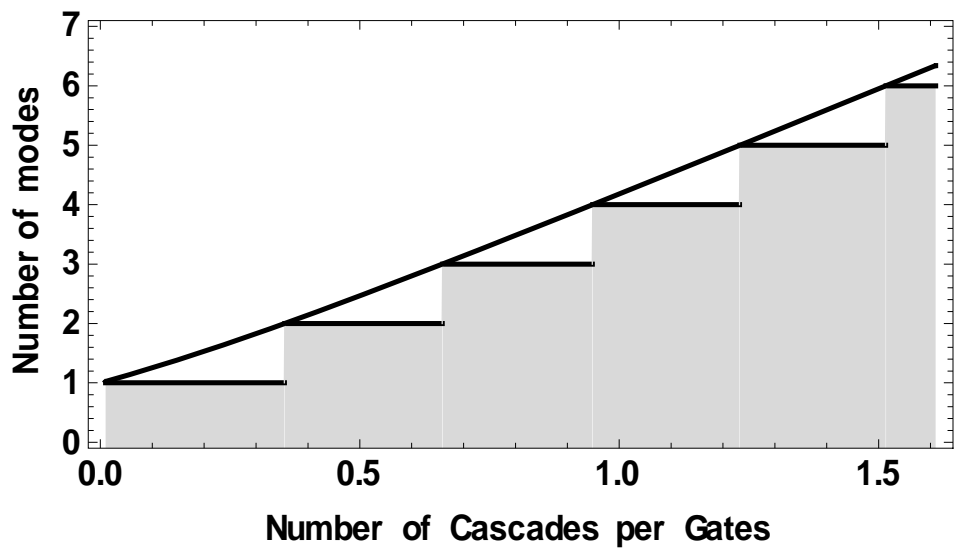

**Figure 6.** Shows the number of modes $M$, calculated from the second equation of Equation (28). The shaded step function is the integer part of the the continuous curve. The coincidence curve in Figure 5 is connected to the present figure according to the simple relation. On the horizontal axis the pump parameter $Nw$ is varied from zero up to the value 1.6.

The agreement with the two theoretical results of completely different origin, and the agreement of both of them with the experimental data is certainly not an accidental numerical coincidence. The quantum mechanical result given by eq. (8) of Ref. [58] (see quotation above) can be related to the number of relevant modes $M$ in a quite straightforward manner. The quantity $\alpha_{QM}$ can be brought to the following equivalent form

$$\alpha_{QM} = 1 - \frac{f^2(w)}{[f(w)+Nw]^2} = 1 - \frac{1}{\overline{M}}, \quad \overline{M} \equiv \frac{[f(w)+Nw]^2}{f^2(w)}. \tag{29}$$

Now let us express $\overline{M}$ in terms of the measured effective transmission and reflection coefficients

$$T_{obs} \equiv \frac{N_{2t}}{N_{2t}+N_{2r}}, \quad R_{obs} \equiv \frac{N_{2r}}{N_{2t}+N_{2r}}, \quad \overline{M} = \frac{[f(w)+Nw]^2}{f^2(w)} = T_{obs} \times R_{obs} \times \frac{(Nw)}{(Nw)_0}. \tag{30}$$

If the pumping is close to zero, then $\overline{M} \approx 1$, which, on the other hand means that the excess coincidence numerically equals to the accidental coincidences. The detailed mathematical derivation will be presented elsewhere [39]. It is interesting to compare the values of $\overline{M}$, calculated directly from the experimental data [58], with the values of our original mode function $M$ on the basis of its analytic form in Equation (28). In Figure 7 we show the transmission and reflection coefficients, their product and the values of $\overline{M}$ for different average number of cascades during a gate. We have calculated these numbers from the published data in Ref. [58]. In the lower right figure we see that $\overline{M}$ is a linear function of the effective overlap, $(Nw)/(Nw)_0$, with tangent 4.

SPACE FOR FIGURE 7

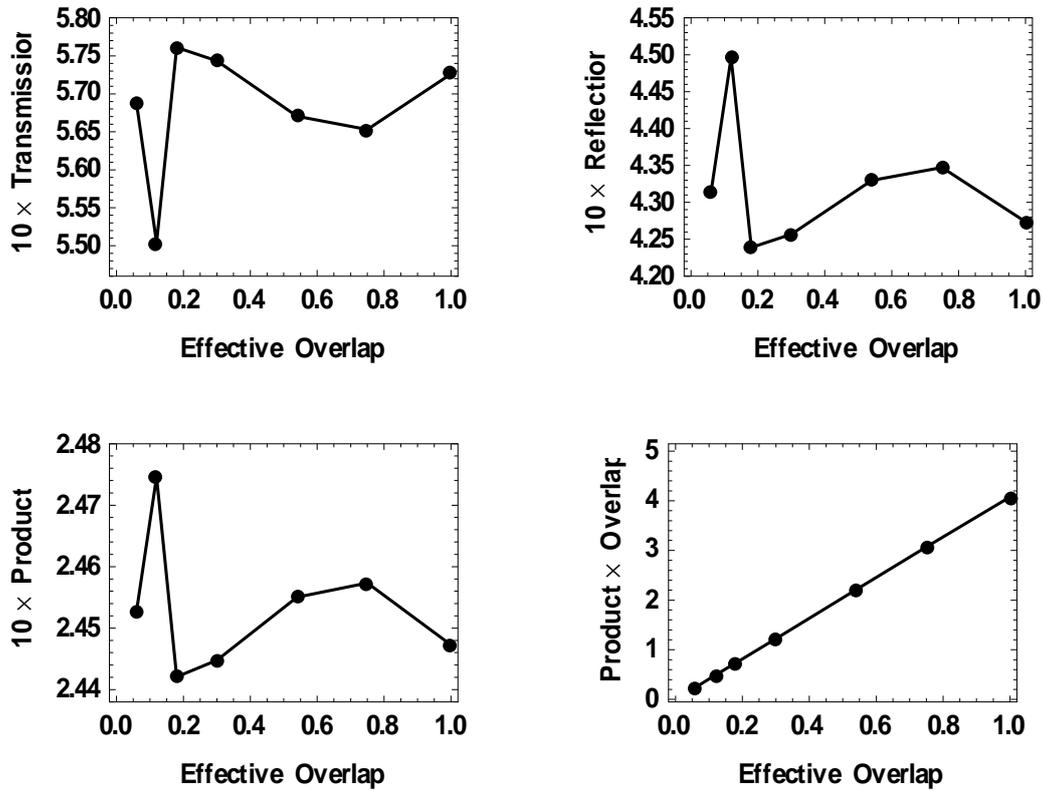

**Figure 7.** These figures have been exclusively drawn on the basis of the experimental data published in Ref. [58], and calculated according to Equation (30). Each figure show the dependencies on the average number of cascades $Nw = \{0.06, 0.12, 0.18, 0.3, 0.54, 0.75, 1\}$ (called 'Effective Overlap' on the abcissas) during a gate. $N$ is the rate of excitation of the cascades, and $w$ is the gate duration. Upper left: empirical transmission coefficient $T_{obs}$. Upper right: empirical reflection coefficient $R_{obs}$. Lower left: the product $T_{obs} \times R_{obs}$. Lower right: the product $T_{obs} \times R_{obs} \times [(Nw)/(Nw)_0]$. Here $(Nw)_0 = 0.06$ was the lowest value of the number of cascades during the gates in the experiments. Each curves, except for the last one, are quite irregular, without displaying any seemingly systematic dependence. However, if we calculate the 'product of the upper two curves', and multiply the result with the relative number of gates, then, without any forced adjustments, we receive an 'ideal' monotonously proceeding rising straight line. In fact, the lower right figure illustrates the illustrates the invariance property $T_{obs} \times R_{obs} \times [1/(Nw)_0] = 4$, which is valid throughout the whole interval $0 \leq Nw \leq 1$.

Figure 8 displays the comparison of the theoretical values $M$ and the empirical values $\overline{M}$. The figure suggests that these two quantities asymptotically coincide, and it is also remarkable that, at least for larger values of the overlap parameter, the measured points are sitting on the steps which represent the integer values of  the analytic formula given by Equation (28).

SPACE FOR FIGURE 8

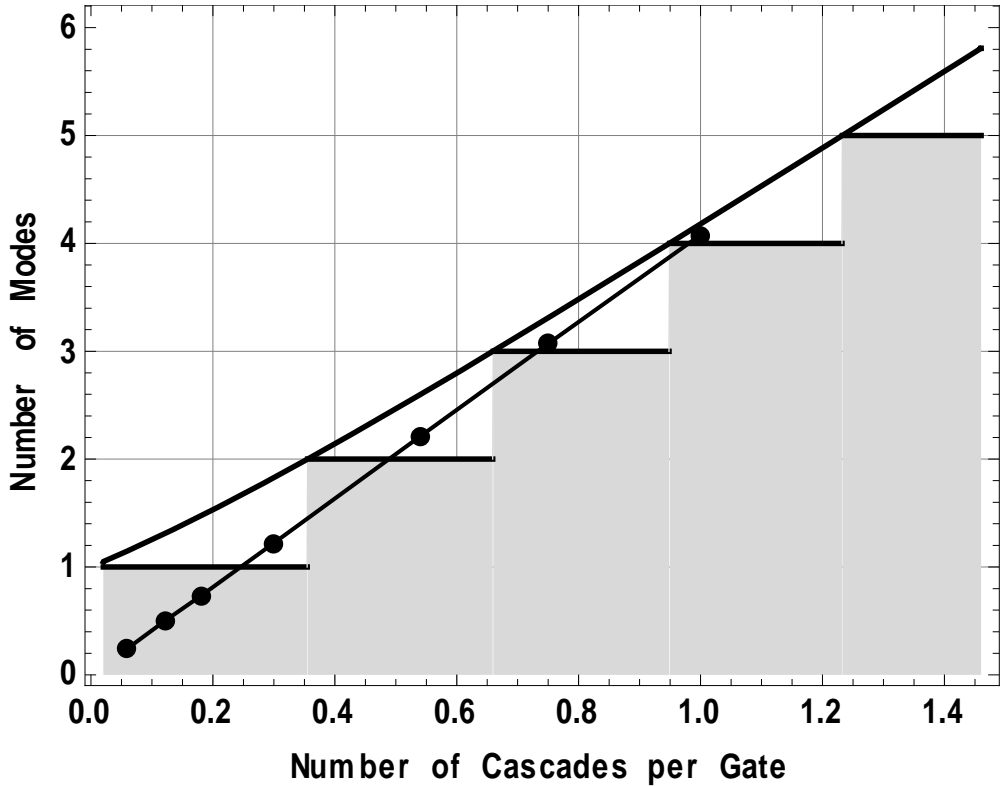

**Figure 8.** Shows the number of modes calculated from the second equation of Equation (28). The shaded step function is the integer part of the the continuous curve, like in Figure 6. The straight line with dots represent the calculated values of the empirical formula we found on the basis of the experimental results published by Aspect and Grangier [58]. The dots correspond to the seven measured values of the number of coincidences shown in Table 1. We emphasize that no fitting has been applied by drawing these curves.

We consider the results embodied in Figure 8 as a consistency check for the correct usage of the concept of modes in the present paper. We think that the semi-phenomenological rule (essentially for the regression line) we have presented in Equation (30), may also give a hint for the appreciation of the physical existence of modes as natural degrees of freedom in quantum theory.

## 7. Conclusions

We have discussed intensity-intensity (or particle-particle) correlations, under the following assumptions. The source of the quanta is considered as an external source (which pumps the measuring apparatus with a prescribed but arbitrary statistics governed by the emission process), that is, no back-action on the source is described. The second condition is that the counting process is linear, which means, that at a given sharp instant of time at most only one quantum may be detected. Because in reality no 100% pumping of any finite entrance port is possible, during the excitation of the apparatus there is always a finite probability that neither of the two detectors are excited. The distribution of the detection events is described by classical probability theory (on the basis of trinomial distributions for the ternary outcomes), but this does not mean at all that for instance the measured electromagnetic radiation is considered as an infinitely divisible continuous entity. Thus the term 'classical' does not refer to some classical (Maxwell) field. In the present description of *counting experiments*, *under the above conditions*, the operator algebra of the quantized field amplitudes (acting on particle number states representing here the removal of the *integer number of quanta*) is replaced by classical probability calculus, on the basis of a simple and clear physical picture. (At this point, we also note that, the present description, however, has nothing to do with the stochastic electrodynamics, operating with continuous field quantities.) The joint distribution of the counting events (forming a Boole algebra − or a $\sigma$-algebra, in general − which is associated to a given experimental situation) are determined by $P(\xi = m, \eta = k)$ of Equation (8) through the statistical weights $W_n$ of the sequences given by Equation (9). The boundary conditions met by the quantum, the response properties of the measuring apparatus, the source and beamsplitter characteristics are all incorporated into these weigths $W_n[M(\boldsymbol{R_1}, T_1; \boldsymbol{R_2}, T_2 \mid S, BS, D)]$, which are governed by the number of relevant modes (number of degrees of freedom) $M$. In the present description the wave character (the periodic continuum character) of the field manifest itself through these generalized mode functions. Under the assumptions stated above, our results (at least in the various examples we have discussed) coincide with the ones derivable from the standard fourth-order normally-ordered quantum coherence functions $\langle E^{(-)}(x_1)E^{(-)}(x_2)E^{(+)}(x_2)E^{(+)}(x_2)\rangle$. In order to consider more general situations – for instance when the source, the propagating quanta and the detector are forming a joint system, or the interaction is nonlinear – the present description must be generalized (e.g. by modifying the Boole algebra). On the other hand, one should also keep in mind that the quantum coherence functions mentioned before, should also be replaced in such cases with more general formulae (as have been *a priori* done by Goldberger, Lewis and Watson [14] in 1963), thus a general comparison of the two descriptions is far not simple and by now it is an unsolved problem. We think that it is a definite advantage of the present description, that the bosons and the fermions are considered on same footing, since at the level of the '*n*–sequences' the same trinomial distribution applies, as has been clearly emphasized in Section 2. Another advantage is the use of the Boole algebra, which conforms ordinary common sense logic, instead of the non-distributive lattice of Hilbert subspaces. At the simplest phenomenological level we have used, we have separated the 'discrete, quantum part' and the 'continuous, wave part' in the linear counting experiments, thus we were able to simply keep track of the algebra of possible counting events.

We have presented a unified treatment of classic and recent Hanbury Brown and Twiss type counting experiments for both bosons and fermions. As in general, in these experiments, too, the removal of the quantum takes place in a the microscopic surroundings of the detector atom, thus even if one counts altogether $10^{12}$ events (e.g. photoelectrons in the *macroscopic* detectors), the relative chance to observe *true* coincidences is on the order of $(10^{12}/10^{23})^2 = 10^{-22}$ for two detectors of lateral size $1cm^2$ which is swept throug by light within $0.3ns$. Typically the total number of counts is on the order of a couple of hundreds of thousands, or perhaps some millions, which is still negligible in comparison with the huge Avogadro number. One may say that the set of spatio-temporal detection points is practically a set of measure zero in the macroscopic apparatus. This assumption is firmly supported by the early quantum mechanical works of Goldberger and Watson [17]. Let us quote them now: "A comparison with such descriptions as those given by Purcell[11] and Twiss and Little[12] may be convenient at this point. They discuss the number of coincidences $N_c$ between particles during the time $T$. This may be done most easily when the expected number of particle arrivals in the resolving-time interval $\Delta\tau_r$ is much less than unity. Then (with a little more attention given to the definition of $\Delta\tau_r$) $\langle N_c \rangle = \Delta\tau_r T \langle G_{12} \rangle$. [ Eq. (2.32) ] The term $\Delta\tau_r T \langle G_1 \rangle \langle G_2 \rangle$ has been described[12] as due to 'random coincidences' and $\Delta\tau_r T \langle G_{12} \rangle$ as due to particle 'clumping'. This kind of description is picturesque, but of limited applicability (as, for example, to the case of electron beams or to the case in which many particles are counted during one resolving time $\Delta\tau_r$)." Here $\langle G_1 \rangle$ and $\langle G_2 \rangle$ are the mean fluxes of counts at detector A and B, respectively. The excess coincidences stem from real joint counts coming from two different gates, or more precisely, from two copies of the spatio-temporal volume of the macroscopic detector. Thus, when the experimentator changes the effective mismatch of these two copies, then, in fact sweeps the coherence volume of one single quantum. The description of this overlap is based on the calculation of the Green's function of the *classical Maxwell fields or de Broglie waves*, which must be matched to the boundary conditions set by the experimenter [62], [63]. This is also a crucial point in the discussion of the very nature of entanglement, e.g. in parametric down-conversion [54], [64], [65], or in the 'ghost-imaging' experiments [66-71]. However, the description of entanglement (which means in the simplest case a *coherent occupation of two orthogonal modes by two quanta* [65]), needs a different Boole algebra, and, moreover a Schmidt-type analysis should likely be performed, concerning the relevant modes

Finally we would like to note the following. Various, and very sophisticated experiments have been performed recently at extremely low intensities of the probed photon beams, in the so called 'single-photon regime', or in the 'few-photon regime'. The *energy elements h*ν occupying the available phase-space cells (modes) in an experiment are *indistinguishable*, of course, like equal fractions of the kinetic or potential energy of a body. According to quantum electrodynamics, photons (in general, bosons and fermions, as energy elements or quanta) are indistiguishable from each other. They are, in particular trivially indistinguishable from *themselves* when they lonely occupy one whole coherence volume, and are absorbed either by one or the other detector. The measured '*mutual indistinguishability*' in the recent correlation experiments relies on the very narrow spectral band of the beams, and this secures that *the well separated consecutive quanta are indistinguishable*. We have shown above in several examples that the number of relevant (not necessarily plane wave) modes can quite simply be estimated in a given experiment by keeping in mind the number of relevant coherence volumes attached to the locations of the absorptions. This picture may perhaps serve as a usable guide for an intuitively clear interpretation of some earlier and recent experiments.

**Acknowledgements.** This work has been supported by the Hungarian National Scientific Research Foundation OTKA, Grant No. K73728. I thank Professor H. Rauch for many valuable discussions, and for bringing my attention to the neutron correlations measured in recent beam splitter experiments. I also thank Professor M. Iannuzzi for providing me with his latest unpublished results on the temporal autocorrelation experiments with neutrons. In addition, I thank the unknown Referees for their constructive criticism and valuable comments, which helped me in compiling the final version of the present paper.